\documentclass[aps,11pt,showkeys,nofootinbib]{revtex4-1}

\usepackage{amsmath,amssymb,amsfonts,amsthm}
\usepackage{amsbsy} 
\usepackage{epsfig}
\usepackage{latexsym}

\usepackage[dvipsnames]{xcolor}
\usepackage{soul} 
\usepackage{subcaption}
\usepackage{braket}

\input amssym.def
\input amssym.tex

\usepackage{hyperref}

\newcommand{\mRe}[1]{\text{Re}(#1)}
\newcommand{\mIm}[1]{\text{Im}(#1)}

\newcommand{\hubb}[0]{\mathcal{H}}
\newcommand{\meqref}[1]{eq.\,(\ref{#1})}

\newcommand{\mf}[0]{\mathcal{F}} 

\newcommand{\mfr}[0]{\mathcal{F}_\rho}  
\newcommand{\mfrr}[0]{\mathcal{F}_{\rho \rho}}

\newcommand{\mr}[0]{\mathcal{R}}

\newcommand{\my}[0]{\mathcal{Y}}

\newcommand{\ma}[0]{\mathcal{A}}

\newcommand{\mc}[0]{\mathcal{C}}

\newcommand{\tPhi}[0]{\Tilde{\Phi}}

\newcommand{\mh}[0]{\mathfrak{H}}

\newcommand{\rt}[0]{\phi}

\newcommand{\pot}[0]{\Tilde{V}(\phi)}

\def\be{\begin{equation}}
\def\ee{\end{equation}}
\def\dd{{\rm d}}
\def\bes{\begin{eqnarray}}
\def\ees{\end{eqnarray}}

\begin{document}

\title{Gauge-invariant perturbations at a quantum gravity bounce}
\author{Steffen Gielen}
\email{s.c.gielen@sheffield.ac.uk}
\author{Lisa Mickel}
\email{lmickel1@sheffield.ac.uk}
\affiliation{School of Mathematics and Statistics, University of Sheffield, Hicks Building, Hounsfield Road, Sheffield S3 7RH, United Kingdom}
\date{\today}

\begin{abstract}
We study the dynamics of gauge-invariant scalar perturbations in cosmological scenarios  with a modified Friedmann equation, such as quantum gravity bouncing cosmologies. We work within a separate universe approximation which captures wavelengths larger than the cosmological horizon; this approximation has been successfully applied to loop quantum cosmology and group field theory. We consider two variables commonly used to characterise scalar perturbations: the curvature perturbation on uniform-density hypersurfaces $\zeta$ and the comoving curvature perturbation $\mathcal{R}\,$. For standard cosmological models in general relativity as well as in loop quantum cosmology, these quantities are conserved and equal on super-horizon scales for adiabatic perturbations. 
Here we show that while these statements can be extended to a more general form of modified Friedmann equations similar to that of loop quantum cosmology, in other cases, such as the simplest group field theory bounce scenario, $\zeta$ is conserved across the bounce whereas $\mathcal{R}$ is not. 
We relate our results to approaches based on a second order equation for a single perturbation variable, such as the Mukhanov--Sasaki equation.
\end{abstract}

\maketitle

\section{Introduction}

On large scales, the observable Universe can be described in simple terms: it is very close to a homogeneous, isotropic (and probably spatially flat, although that is less clear \cite{closedFLRW, closedFLRW2}) Friedmann--Lema\^{i}tre--Robertson--Walker (FLRW) universe, with nearly scale-invariant small scalar perturbations. The challenge for all approaches to modern cosmology is to find an explanation for this observed structure, and to resolve various puzzles inherent to the currently accepted $\Lambda$CDM model of cosmology. This challenge is often seen as an opportunity for theories of quantum gravity to connect to observations, in particular since the $\Lambda$CDM model features the Big Bang singularity which signals its own fundamental incompleteness. The relative simplicity of our Universe means that one does not need to understand quantum gravity in full generality to say something about cosmology: all that is needed is a formalism powerful enough to deal with approximately homogeneous and isotropic universes. Various quantum-gravity inspired cosmological scenarios can indeed describe the evolution of perturbations across a quantum bounce \cite{BounceReview}.

The description of cosmological perturbations in the standard framework is based on gauge-invariant perturbation variables \cite{cosmopert}, characterised by their invariance under infinitesimal diffeomorphisms, which correspond to physical perturbations.
On top of this notion of gauge invariance, there are gauge-invariant perturbation variables with a particularly direct physical interpretation.
The {\em curvature perturbation on uniform-density hypersurfaces} $\zeta$ is directly related to cosmological observations at late times, in particular of the cosmic microwave background (CMB) \cite{BounceReview}. 
The notion of what characterises a `good' cosmological perturbation variable may not extend straightforwardly to quantum gravity, where the notions of gauge invariance and diffeomorphisms may be modified \cite{BojowaldPaily}, or where one may not have access to an effective action for perturbations. The physical mechanism for the generation of cosmological perturbations may also be different from the most commonly assumed framework of inflation, where they arise from quantum fluctuations on an effectively classical background. It is important to be clear about which assumptions from standard perturbation theory are carried over to a particular quantum gravity formalism of interest. \\

In this paper we study gauge-invariant scalar perturbations in quantum gravity bounce scenarios governed by a modified Friedmann equation which reduces to general relativity at low energies but includes high-curvature corrections. This starting point is familiar from the standard effective dynamics of Loop Quantum Cosmology (LQC) \cite{LQC, LQC_2} but clearly more general\footnote{As a basic example, generalisations of LQC can already lead to more complicated Friedmann equations \cite{GLQCnew}.}. 
We do not assume an effective spacetime description of perturbations, but ask what properties of the dynamics of perturbations can be obtained from the Friedmann equation and homogeneous matter dynamics alone. 
Consequently, we work in the separate universe picture for long-wavelength cosmological perturbations \cite{SepUniv, SepUniv_2, hamSU}, which suggests that such perturbations can be well approximated by a universe consisting of many independent, locally homogeneous patches, each governed by a local Friedmann equation and dynamical equations for matter. 
The separate universe approach has already been studied in LQC \cite{LQCsepUniv} and Group Field Theory (GFT) \cite{GFTsepUniv}. Our work extends the results of \cite{LQCsepUniv} to more general modified Friedmann equations such as those appearing in GFT, while also going beyond \cite{GFTsepUniv} in that we study gauge-invariant perturbation variables.

In standard cosmology the quantity $\zeta$ is conserved on super-horizon scales for adiabatic perturbations, $\zeta'=0$ (see, e.g., \cite{conserved}; here and in the following $'$ refers to derivative with respect to an arbitrary time coordinate). 
This property is very important: it means one only has to follow the evolution of cosmological perturbations while they are within the Hubble horizon. 
Typically, in a cosmological scenario that aims to solve the horizon problem, perturbations are initially generated deep inside the Hubble horizon, then leave the horizon as the Universe expands or contracts, only to re-enter later as amplified, classical perturbations. 
In bounce scenarios, the point of exiting the horizon is often in a contracting phase before the bounce and re-entry happens in the subsequent expansion phase \cite{BounceReview}. (A subtle point which we will get back to later is that at the bounce itself the horizon is necessarily infinite so {\em all} modes are sub-horizon.) 
We review the derivation of the conservation law for $\zeta$  from which one can readily see that it will continue to hold for adiabatic perturbations in quantum gravity scenarios of interest, where the continuity equation remains unaltered. 
 We also study a related quantity, the {\em comoving curvature perturbation} $\mathcal{R}$. This quantity usually satisfies a similar conservation law for long-wavelength modes, $\mathcal{R}'=0$, and in general relativity one can show that $-\zeta=\mathcal{R}$ on super-horizon scales (with appropriate sign conventions). This equality is no longer guaranteed if one goes beyond general relativity as we do here. 
Indeed the simplest GFT bounce scenario gives an example of quantum-gravity inspired cosmological dynamics for which $\zeta$ is still conserved, but $\mr$ is not.

This article is organised as follows: In sec.\,\ref{sec:cosmPert}, we give a brief introduction to standard cosmological perturbation theory, including the definitions of $\zeta$ and $\mr$, and introduce the separate universe framework. 
We then proceed to rederive the well-known conservation law for $\zeta$ in the case of adiabatic perturbations in sec.\,\ref{sec:consZeta}. The main results of this paper are contained in sec.\,\ref{sec:genPert}: we first derive perturbation equations for long-wavelength modes starting from a general modified Friedmann equation without specifying a lapse or choosing a specific gauge for perturbation variables. We then comment on the meaning of different gauge choices in the separate universe framework (sec.\,\ref{sec:gauge}). Working in the comoving gauge, we  study a class of modified Friedmann equations, where the modification is a function of the energy density $\rho$ only (a special case of which is LQC), in sec.\,\ref{sec:mfRho}. 
We find that for this particular case $\mr' = \zeta' = 0$ continues to hold.
In sec.\,\ref{sec:GFTexample} we consider a modified Friedmann equation obtained from GFT as an example for which $\mr'\neq0$.
Our main results are based on solving first-order equations of motion, but in sec.\,\ref{sec:MS} we compare this strategy to approaches based on a single second order equation of motion for a single perturbation variable, such as the Mukhanov--Sasaki equation.
We finally conclude in sec.\,\ref{sec:conc}.

\section{Cosmological perturbations}\label{sec:cosmPert}

The usual starting point in cosmological perturbation theory is to model the Universe as a flat FLRW universe with inhomogeneous perturbations.
We will follow this assumption even though the observational status of spatial curvature is not fully settled. This is because we are interested in the behaviour of perturbations near  a bounce, where spatial curvature would be subdominant, and because the quantum gravity scenarios we are interested in might prefer flat FLRW geometries (see, e.g., \cite{GFTscalarcosmo, GFTscalarcosmo_2} for the situation in GFT cosmology).
The general form of the perturbed line element at linear order for scalar perturbations, following standard conventions \cite{cosmopert,BaumannNotes}, is \footnote{The tilde over $\tilde\Phi$ is used to distinguish this lapse perturbation variable (which is not gauge-invariant) from its gauge-invariant analogue, the Bardeen variable $\Phi$.}
\bes
\dd s^2 &=& -N^2(t)\left(1+2\tilde\Phi(t,x^i)\right)\dd t^2 + 2N(t)a(t)\,\partial_i B(t,x^i)\,\dd t\,\dd x^i \nonumber
\\&&+ a^2(t)\left[\left(1-2\psi(t,x^i)\right)\delta_{ij}+2\partial_i\partial_j E(t,x^i)\right]\dd x^i\,\dd x^j\,.
\label{pertmetric}
\ees
For concreteness we have added the functional dependence of all variables explicitly: we have the background scale factor $a(t)$ and lapse function $N(t)$ (which only depend on time) and perturbations $\tilde\Phi,\psi,B$ and $E$, which in general depend on space and time. 

In cosmological bounce scenarios, departures from general relativity are expected to become relevant near the bounce, 
where most modes are super-horizon (as mentioned earlier, at the bounce itself all modes are sub-horizon).
We will therefore only be interested in a long-wavelength approximation in which spatial gradients are neglected, so that 
\be
\tilde\Phi(t,x^i)\rightarrow\tilde\Phi(t)\,,\quad \psi(t,x^i)\rightarrow\psi(t)\,,\quad \partial_i B\rightarrow 0\,,\quad \partial_i\partial_j E\rightarrow 0\,.
\label{longwavel}
\ee
$\tilde\Phi$ and $\psi$ can then be seen as perturbations in the background quantities $N$ and $a$, which is the essence of the separate-universe idea: one can consider a universe composed of many independent, locally homogeneous flat FLRW patches, with local lapse and scale factor
\be
N_{{\rm loc}} = N(1 + \tilde\Phi)\,,\quad a_{{\rm loc}} = a(1 - \psi)\,,
\label{flrwperturb}
\ee 
where $N$ and $a$ are now considered as averages over many patches, and $\tilde\Phi$ and $\psi$ as small quantities ($O(\epsilon)$ with $\epsilon\ll 1$) characterising the difference between one patch and the average. We see that the metric for each patch would then be given by (\ref{pertmetric}) with (\ref{longwavel}) (here and throughout the paper, quantities of quadratic and higher order in perturbation variables are $O(\epsilon^2)$ and will be dropped, as we work within linear perturbation theory). 
The separate universe picture is particularly useful in quantum gravity scenarios that do not allow for a satisfactory description of inhomogeneities.

If matter is described by an energy density $\rho$ and pressure $P$, one can analogously introduce perturbation variables $\delta\rho$ and $\delta P$ with
\be
\rho_{{\rm loc}} = \rho + \delta\rho\,,\quad P_{{\rm loc}} = P+\delta P\,.
\label{matterperturb}
\ee 
In the case where the matter content of the very early  universe is given by a single scalar field $\phi$ with potential $\pot$ (such as in inflation, LQC or GFT), 
the energy density and pressure are given by 
\begin{align}
\rho = \frac{\phi'^2}{2N^2} + \pot \, , \qquad P = \frac{\phi'^2}{2N^2} - \pot
\label{eq:rhoPscalarField}
\end{align}
and the dynamics of the scalar field are determined by the Klein--Gordon equation \begin{align}
\phi'' - \frac{N'}{N} \phi' + N^2 \frac{\dd \pot}{\dd \phi} + 3 \mh \phi'  = 0\, ,
\label{eq:klGordon}
\end{align}
where we define $\mathfrak{H}=a'/a$ (which reduces to the usual Hubble parameter $H$ if $t$ is chosen to be proper time).
For $\phi_{{\rm loc}}=\phi+\delta\phi$ one obtains the perturbations for the energy density and pressure by perturbing \eqref{eq:rhoPscalarField} at linear order:
\begin{align}
\delta \rho = \frac{\phi'^2}{ N^2} \left(\frac{\delta \phi'}{\phi'} - \tPhi\right) +\frac{\dd \pot}{\dd \phi} \delta \phi\, , \qquad \delta P = \frac{\phi'^2}{N^2} \left(\frac{\delta \phi'}{\phi'} - \tPhi\right) -\frac{\dd \pot}{\dd \phi} \delta \phi\,.
\label{eq:pertRhoP}
\end{align}

In cosmological perturbation theory, there exist two notions of gauge invariance,
 one related to the choice of the lapse $N$ and another to the choice of the coordinate system of perturbations. 
The first allows an arbitrary change of the background time coordinate $t\to f(t)$; the second encodes the invariance under \emph{infinitesimal} diffeomorphisms  $x^\mu \to x^\mu + \xi^\mu$, where $\xi^\mu$ is $O(\epsilon)$.
 It is the second gauge freedom we refer to when we discuss gauge invariance and different gauge choices in the following, keeping in mind that our separate universe approximation will reduce the relevant gauge transformations to those of the form $t \to t+\xi^0(t)$. For more details, see \cite{cosmopert, BaumannNotes}.
The metric and matter perturbation variables in \eqref{pertmetric} are not gauge-invariant, but have certain transformation properties under infinitesimal diffeomorphisms, such that gauge-invariant variables need to be obtained by combining perturbation variables in a suitable manner. 
The two gauge-invariant perturbation variables we have mentioned above are defined by (where different sign conventions for $\zeta$ exist in the literature and we follow the one in \cite{conserved} and going back to \cite{BST1983})
\be
-\zeta = \psi + \frac{\mathfrak{H}}{\rho'}\delta\rho\,,\quad \mathcal{R} = \psi+\frac{\mathfrak{H}}{\phi'}\delta\phi\,,
\label{gaugeinvperts}
\ee
where the second definition specifically refers to a scalar field. 
As such, the variables in \eqref{gaugeinvperts} are invariant under infinitesimal diffeomorphisms acting on the metric and matter fields at $O(\epsilon)$ \cite{cosmopert}. Importantly, we will assume that {\em the same notion of gauge invariance applies in our quantum gravity scenarios of interest} so that (\ref{gaugeinvperts}) are gauge-invariant, and hence potentially observable, also for our modified gravitational dynamics. This is certainly an assumption given that the action of diffeomorphisms might receive quantum corrections as in LQC \cite{LQCanomaly}, but in the absence of a full spacetime picture we need to make such an assumption to proceed. Since we are only interested in long-wavelength perturbations, we only need to assume that diffeomorphisms act as in general relativity in that limit, which is true in LQC and might be seen as an admissible assumption to make: naively, the longest wavelengths should be least sensitive to any quantum gravity corrections.

Within general relativity  coupled to a scalar field, one can show that $-\zeta=\mathcal{R}+O(k^2)$, where $k$ is the wavenumber in a Fourier decomposition, so that in the separate universe limit $k\rightarrow 0$ 
(or, more accurately, a limit in which the physical wavelength of a mode is much larger than the Hubble horizon: $k\ll\frac{a'}{N}$)
 we have $-\zeta=\mathcal{R}$. 
For this reason the two quantities are often treated as interchangeable when long-wavelength modes are studied, but again it is not clear whether a similar relation holds in more general cosmological models of the type we are interested in.
We would like to point out that during slow-roll inflation one can approximate
\be
\rho=\frac{\phi'^2}{2N^2}+\pot\approx \pot\quad\Rightarrow \;\rho'\approx \phi' \frac{\dd \pot}{\dd \phi}
\ee
and hence in this case $-\zeta\approx\mathcal{R}$ without using any gravitational field equations. We will not be interested in slow-roll inflation, but consider scalar fields with general potentials such that the high-energy regime can be dominated by kinetic energy.\\

\section{Simple conservation law for $\zeta$}\label{sec:consZeta}
We start by deriving the simplest conservation law for the perturbation variables we are considering: $\zeta$ is conserved on large scales for adiabatic matter perturbations if the continuity equation for matter is unchanged. 
This is an old result in cosmology \cite{conserved} which extends directly to many quantum bounce scenarios, in particular to LQC \cite{LQC, LQC_2} and GFT \cite{GFTscalarcosmo, GFTscalarcosmo_2,deparamcosmo} which do not introduce any alterations to the dynamics of the matter content of the universe.
Rederiving this result illustrates the philosophy behind the separate-universe approach.

The continuity equation satisfied by the background variables reads
\be
\rho' + 3 \mathfrak{H}(\rho+P)=0\,,
\label{continuity}
\ee
and if we introduce locally perturbed variables according to (\ref{flrwperturb}) and (\ref{matterperturb}), assuming that the continuity equation also holds in each local patch, we find the perturbed continuity equation
\be
\delta\rho' + 3 \mathfrak{H}(\delta\rho+\delta P)-3\psi'(\rho+P)=0\, ,
\label{pertcontinuity}
\ee
using that $\mathfrak{H}_{{\rm loc}}=\mathfrak{H}-\psi'$. Hence,
\be
-\zeta'=\psi' + \left(\frac{\mathfrak{H}}{\rho'}\delta\rho\right)' = -\left(\frac{1}{3(\rho+P)}\right)'\delta\rho+\frac{\mathfrak{H}(\delta\rho+\delta P)}{\rho+P}= \frac{\rho'+P'}{3(\rho+P)^2}\delta\rho+\frac{\mathfrak{H}(\delta\rho+\delta P)}{\rho+P}
\ee
using (\ref{continuity}) and (\ref{pertcontinuity}). The assumption of adiabatic perturbations means that we can write $\frac{\delta P}{\delta\rho}=\frac{P'}{\rho'}$ and one finally obtains, after again using (\ref{continuity}), that $\zeta'=0$ for these perturbations. 
This argument relies on the long-wavelength limit, since the perturbed continuity equation \eqref{pertcontinuity} would in general contain terms involving spatial derivatives (sec.\,\ref{sec:ddpsi}). It does however not use the gravitational dynamics, and hence holds in many scenarios beyond general relativity.

One commonly introduces the equation of state parameter $w = \frac{P}{\rho}$ and the sound speed $c_s^2 = \frac{P'}{\rho'}$ and using these definitions together with the continuity equation (\ref{continuity}), one finds that
\be
w' = -3 \mathfrak{H}  (w +1) (c_s^2 -w )\,.
\label{eq:dw}
\ee
In the case of a perfect fluid, $w$ is constant, $c_s^2 = w$, and the adiabaticity condition is always satisfied. 
A scalar field can mimic a perfect fluid if the potential is chosen such that it has a constant equation of state parameter (at least for the time scales one is interested in). In particular, this is the case for a massless scalar field, where $w=1$. 
For more general scalar field dynamics, there is exchange between kinetic and potential energy in the scalar field and perturbations can not generally be assumed to be adiabatic; however, in general relativity a single scalar field can generally only produce adiabatic perturbations on large scales, so that $\zeta'=0$  \cite{GRscalarAdiabatic}. As we will show in sec.\,\ref{sec:mfRho}, $\zeta'=0$ still holds for a single scalar field for specific forms of modified gravitational dynamics.
However, for general modifications to the Friedmann equation, one cannot conclude that a single scalar field with an arbitrary potential induces only adiabatic perturbations.

\section{Generalised Friedmann dynamics and their perturbations}\label{sec:genPert}

We now introduce gravitational dynamics given by generalised Friedmann equations of the type expected in many approaches to quantum gravity. For this generalised Friedmann equation, we write
\be
\frac{\mathfrak{H}^2}{N^2}=\frac{\kappa}{3}\rho\,\mathcal{F}\, ,
\label{eq:genFried}
\ee
where $N$ is a general choice of lapse, $\kappa=8\pi G$ is a rescaled Newton's constant and the function $\mathcal{F}$ encodes the quantum gravity corrections to the Friedmann equation of general relativity. ($\mf=1$ then corresponds to the general relativistic Friedmann equation.)

The perturbation equations that follow in this section are independent of the specific form of $\mf$, but we would like to give two examples that we will get back to later, namely the modified Friedmann equations of LQC and GFT.
As we will see, these two examples characterise two qualitatively different cases, where $\mathcal{F}$ and its perturbations either depend only on $\rho$ and $\delta \rho$ (sec.\,\ref{sec:mfRho}) or on other variables as well (sec.\,\ref{sec:GFTexample}). 
When deriving the effective Friedmann equation in LQC or GFT one assumes that the matter content of the universe is given by a single massless scalar field, such that the energy density is given by $\rho = \frac{\pi_\phi^2}{ (2 a^6)}$, where the scalar field momentum $\pi_\phi$ is a constant of motion. 
In the standard effective dynamics of LQC \cite{effectiveLQC}
\be
\mathcal{F}_{\rm LQC}=1-\frac{\rho}{\rho_{{\rm c}}}\,,
\label{eq:mfLQC}
\ee
where $\rho_{{\rm c}}$ is a universal, maximal energy density, which characterises the regime in which quantum gravity corrections become relevant. 
For phenomenological applications $\mf_{\rm LQC}$ is sometimes assumed to hold also for massive scalar fields, even if this cannot be directly derived from the quantum theory (see, e.g., \cite{LQCsepUniv, LQC_genRho, LQC_matterBounce}).
In GFT more general forms appear such as \cite{deparamcosmo}
\be
\mathcal{F}_{\rm GFT}=1+\frac{v_0}{a^3}+\frac{\my}{a^6}\,.
\label{eq:mfGFT}
\ee
Here, $v_0$ is a fixed constant that has the unit of volume and gets its interpretation from the underlying quantum theory. $\my$ on the other hand is a constant of motion rather than a fundamental parameter and so in the separate universe picture will vary from one patch to another, $\delta \my \neq 0$. Its value is related to the volume of the universe at the bounce.

To obtain the dynamics of perturbations for a modified Friedmann equation while being agnostic about the details of the underlying gravitational theory, we proceed as follows. 
An equation of motion for $\mh$ is obtained from the time derivative of \eqref{eq:genFried}\footnote{The equation is written in this form for convenience, however, division by $\mh$ and $\mf$ is not defined at the bounce.},
\begin{align}
\frac{ \mh'}{\mh}  = \frac{ N'}{N} + \frac{1}{2}\left( \frac{ \rho'}{\rho} + \frac{ \mf'}{\mf} \right)\,,
\label{eq:dtauH}
\end{align}
and the first and second order equations of motion for the metric perturbation $\psi$ are obtained by perturbing \eqref{eq:genFried}  and \eqref{eq:dtauH} at linear order (or equivalently, perturbing \eqref{eq:genFried} and then taking the time derivative). Giving different equivalent forms of these equations, we have
\begin{align}
 \begin{split}
 \mh  \psi'  = & -\mh^2\left(\tPhi  +\frac{\delta\rho}{2\rho}+\frac{\delta \mf}{2\mf}\right)\\
  = &  -\frac{\kappa}{3} N^2\left(\rho\mf\,\tPhi  +\frac{1}{2}\left(\mf \delta \rho + \rho \delta \mf\right)\right)\, ,
 \label{eq:dPsi}
   \end{split}\\
 \begin{split}
- \psi'' = & \left(-\frac{N'}{N}-\frac{\mf'}{2\mf}+3\mh\frac{\rho+P}{\rho}\right)\psi' + \mh\,\tPhi' + \frac{\mh}{2}\left(\frac{\delta \mf'}{\mf}-\frac{\mf'}{\mf^2}\delta \mf\right)\\
 & + \frac{\kappa}{2}N^2 \mf (\rho+P) \left(\frac{\delta\rho}{\rho}-\frac{\delta\rho + \delta P}{\rho + P}\right)\\ 
= & - \frac{N'}{N} \psi'  + \mh\,\tPhi'
-  \frac{\kappa}{2} N^2 \mathcal{F} (P+\rho ) \left(  \frac{( \delta P +\delta \rho )}{P+\rho}+  \frac{\delta  \mathcal{F}}{\mathcal{F}}+ 2 \tPhi \right)\\
& + \frac{\mh}{2}\frac{\mf'}{\mf}\left(-\frac{   \delta \mathcal{F}   }{2
   \mathcal{F}}+\frac{\delta \rho }{2 \rho} + \tPhi+\frac{\delta
   \mathcal{F}' }{ \mathcal{F}'} \right)\, .\label{eq:ddPsi}
   \end{split}
\end{align}
These forms can be transformed into each other by using the Friedmann equation (\ref{eq:genFried}) and by using (\ref{eq:dPsi}) to rewrite terms proportional to $\psi'$ in (\ref{eq:ddPsi}).
For $\mf  =1$ (and hence $\mf'=0\,, \ \delta \mf =0 \,, \ \delta \mf' = 0$) the above reduce to the standard general relativistic equations of motions for perturbations in the long-wavelength limit (spatial gradients vanish) as obtained from the Einstein field equations.

So far, we have not assumed a specific form of matter content. Neither have we chosen a specific form of lapse, nor made a gauge choice for perturbation variables. 
A popular choice of time coordinate is conformal time, where $N=a$, and certain gauge choices simplify the perturbation equations further. 
We comment on the question of gauge in the following subsection, and then discuss a specific class of functions $\mf$ for which the perturbation equations above simplify in an LQC-like fashion.

\subsection{Gauge choices in the separate universe approximation}\label{sec:gauge}

Even though one is ultimately interested in the dynamics of gauge-invariant quantities, it is often useful to carry out calculations in a specific gauge.
A popular gauge choice in cosmology is the Newtonian or longitudinal gauge, in which all anisotropic perturbations of the metric tensor are set to zero ($B=E=0$). 
This gauge was also used in previous studies of the separate universe framework  in LQC \cite{LQCsepUniv} and GFT \cite{GFTsepUniv}.
Here we will instead work in the comoving gauge, which for scalar matter is defined by $\delta \phi = 0$ and $E=0\,$.\footnote{There exist different conventions for the comoving gauge: in \cite{ Giesel_2018} the comoving gauge is defined by $\delta \phi =0$ and $B=0$ instead. The above convention is used in, e.g., \cite{BaumannNotes}.}
(As the matter content in LQC and GFT is taken to be a scalar field, we focus on this case in the following.)
For scalar matter the lapse perturbation is directly related to the perturbation of the energy density and pressure \eqref{eq:pertRhoP} in comoving gauge, 
\begin{align}
\delta \rho = -\frac{\phi'^2}{N^2}  \tPhi = - (\rho + P)\tPhi =\delta P \,
\label{eq:tPhiCom}
 \end{align}
and one can write, using (\ref{eq:tPhiCom}) and the continuity equation \eqref{continuity},
\begin{align}
-\zeta = \mr + \mh \frac{\delta \rho}{\rho'} = \mr + \frac{\tPhi}{3}\,.
\label{eq:zetaMr}
\end{align}

The reasons we choose the comoving gauge are threefold. Firstly, the comoving curvature perturbation takes a particularly simple form, $\mr = \psi$. Secondly, when working with relational settings such as GFT, where the scalar field takes the role of a physical clock \cite{GFTscalarcosmo, GFThamiltonian}, the comoving gauge is simply the statement that at an instant of time all patches of the separate universe picture have the same clock value. 
The third reason is more subtle and is connected to the  application of gauge choices in general cosmological perturbation theory to the separate universe picture. 
As already pointed out in \cite{hamSU}, fixing the Newtonian gauge $E=B=0$ does not provide an additional prescription in the separate universe picture, as anisotropic degrees of freedom are already absent in this approximation (see \eqref{longwavel}). 
In general relativity, a relation between the metric and the lapse perturbation in Newtonian gauge is obtained from the off-diagonal spatial components of the perturbed Einstein field equations ($\delta G_{i\neq j} = \kappa \delta T_{i\neq j}$) \cite{cosmopert}, which for $E=B=0$ reduce to
\begin{align}
\partial_i \partial_j \psi - \partial_i \partial_j \tPhi = 0\,.
\label{eq:EFEij}
\end{align}  
They imply that $\psi = \tPhi + x^i h_i(t) + g(t)\,$, where $h$ and $g$ are arbitrary homogeneous functions.
One then usually sets  $\tPhi = \psi$ arguing that perturbations should average out to zero ($\int \dd^3 x\;\psi = 0 = \int \dd^3 x\;\tPhi $) since any homogeneous contribution to the perturbations could be absorbed in the background \cite{cosmopert}. This argument would forbid any nontrivial $h_i(t)$ or $g(t)\,$. In the separate universe framework all spatial gradients automatically vanish and the off-diagonal components of the Einstein field equations are trivially satisfied, so there is no analogue of \eqref{eq:EFEij}.
Equivalently, requiring that $\psi = \tPhi + g(t)$ is not a constraint as the perturbations $\psi, \, \tPhi$ are in any case only functions of time in the separate universe picture. 
 Fixing $\tPhi = \psi$ for the Newtonian gauge in the separate universe framework is therefore an additional assumption, somewhat harder to justify than in usual cosmological perturbation theory. This issue is also discussed in \cite{0801Wands}, where the authors introduce a `pseudo-longitudinal gauge', which ensures $\psi = \tPhi$ throughout as long as it is assumed to hold in some limit.\footnote{
It is also discussed for the Hamiltonian framework in \cite{hamSU}. 
In the Hamiltonian picture the relation $\psi = \tPhi$ cannot be recovered due to the absence of the diffeomorphism constraint in the separate universe framework. 
In \cite{hamSU} the authors recover $\Tilde{\Phi} = \psi$ by redefining the Newtonian gauge, where the redefinition relies on a relation between perturbation variables obtained from the diffeomorphism constraint.} \\

\subsection{A special case: $\mf = \mf(\rho)$}\label{sec:mfRho}
In LQC, the general perturbation equations take a particularly simple form due to the specific form of $\mf$ \eqref{eq:mfLQC}.
In this section we generalise the LQC case by considering a restricted class of corrected Friedmann equations, namely those in which $\mf$ is a function of the energy density $\rho$ only. 
In particular, this means that the perturbation of $\mf$ is proportional to the perturbation of the energy density, $\delta\mf = \frac{\dd \mf}{\dd \rho}\delta \rho$. Unlike the case of LQC, the GFT correction given in \eqref{eq:mfGFT} does \emph{not} fall in this category, $\mf_{\rm GFT} \neq \mf(\rho)$, as $\my$ is perturbed as well.

If we define the quantities
\begin{align}
\mfr  := &  \frac{\dd \mf}{\dd \rho}\,, \qquad \mfrr:= \frac{\dd^2 \mf}{\dd \rho^2}\,, \qquad
\ma  :=  \mf + \mfr\, \rho\,, \label{eq:maDef}
\end{align}
we obtain the following relations for quantities derived from $\mf$, using again the continuity equation \eqref{continuity}:
\begin{align}
\begin{split}
\delta \mf   = & \,\mfr \delta \rho\,, \qquad \mf'  =  - 3 \mh (\rho + P)\mfr\,, \qquad 
\ma' =  - 3 \mh (\rho + P) (2\mfr + \rho \mfrr )\,,
 \\
\frac{\delta \mf'}{\mf'}  = & \, \frac{\mfrr \delta \rho}{\mfr}+\frac{\delta\rho + \delta P}{\rho + P}-\frac{\psi'}{\mh}\,.
\label{eq:mfRhoSimp}
\end{split}
\end{align}
We can then write the second Friedmann equation (\ref{eq:dtauH}) and generalised perturbation equations \eqref{eq:dPsi}-\eqref{eq:ddPsi} for the $\mf(\rho)$ class of modified Friedmann equations as
\begin{align}
 \mh' - \frac{ N'}{N}\mh  & =  - \frac{\kappa}{2}N^2 (\rho + P)\ma\,, \label{eq:mh'SU_Frho}\\
 \mh  \psi' & = -\mh^2\tPhi  - \frac{\kappa}{6} N^2  \ma \,\delta \rho\,,
 \label{eq:psi'SU_Frho}\\
\begin{split}
-\psi'' & = -\frac{N'}{N}\psi'+ \mh \tPhi'  -\frac{\kappa}{2}N^2\left(\ma\,\delta P + 2(\rho+P)\ma\,\tPhi + (\mf+\rho(\rho+P)\mfrr+(2P+3\rho)\mfr)\delta\rho\right) \\
 & = -\frac{N'}{N}\psi'+ \mh \tPhi'  -\kappa \phi' \ma\, \delta \phi'  -  \left(\mfr + \frac{\rho}{2} \mfrr\right)\kappa \phi'^2\, \delta \rho \,,
 \label{eq:psi''SU_Frho}
 \end{split}
\end{align}
where in the last line we used the fact that matter is given by by a scalar field with $\rho+P=\frac{\phi'^2}{N^2}$ (the other equations are general and hold for any matter content). 
Equations \eqref{eq:mh'SU_Frho}-\eqref{eq:psi''SU_Frho} hold in any gauge.
They correspond to the LQC equations reported in \cite{LQCsepUniv} (which are given in conformal time $N=a$, and for a gauge in which $\psi=\tPhi$) for $\ma = 1 - 2\frac{\rho}{\rho_{{\rm c}}}$ and $\mfr=-\frac{1}{\rho_{{\rm c}}}$.

To make further progress, another equation is needed. 
In general relativity, this is the {\em diffeomorphism constraint} arising from
the mixed time-space components of the perturbed Einstein field equations, $\delta G^0_i = \kappa \delta T^0_i$, or \cite{cosmopert}
\begin{align}
\partial_i \left(\mh \tPhi +\psi' - \frac{\kappa}{2} \phi'\, \delta \phi \right) =: \partial_i D & = 0\,,
 \label{eq:0iconstraint}
\end{align}
which implies $D = s(t)$ where $s$ is an arbitrary homogeneous function. 
In the usual formalism, one then argues that perturbations are inhomogeneous functions over a homogeneous background, and any homogeneous contribution to $D$ can be absorbed in the homogeneous background dynamics to justify the requirement that the perturbation variables satisfy $D = 0$. This discussion is analogous to the gauge choice $\psi=\tPhi$ in Newtonian gauge that we discussed in sec.\,\ref{sec:gauge}.

As in the discussion in sec.\,\ref{sec:gauge}, there is no analogue of \eqref{eq:0iconstraint} in the separate universe framework due to the absence of spatial gradients. 
However, in scenarios of interest such as bouncing cosmologies, one expects to set initial conditions in a regime where general relativity holds and one can consider the case where spatial gradients are small, but not exactly zero. Then, at some initial time $t_0$ where initial conditions are set, which can be any time at which the strict $k \to 0$ limit has not been applied yet, \eqref{eq:0iconstraint} implies $D=0$ for the modes of interest. If one can then show that in general $D'(t) =0$ together with the initial condition $D(t_0) = 0\,$, $D=0$ holds throughout the evolution. This means we obtain another effective constraint equation for perturbations which, as we will see below, can be used to infer conservation laws for gauge-invariant perturbation variables.
For a setting with a modified Friedmann equation, \eqref{eq:0iconstraint} represents the low-curvature limit ($\mf \to 1$) of a possibly modified form of $D$. The modified form for $D$ cannot be derived directly but guessed and justified in hindsight, as was done for LQC in \cite{LQCsepUniv}. 
In the following, we introduce a differential equation for suitable $D$ in comoving gauge  and show that it holds for any form of $\mf(\rho)\,$.
We work in the comoving gauge, but a similar calculation can be carried out in the Newtonian gauge (see app.\,\ref{app:constrNewton}). 

Inspired by \cite{LQCsepUniv},  where the corrections to $D$ for a modified Friedmann equation appear in the $\delta \phi$ term only, we assume that in comoving gauge $D$ takes the same form as in general relativity,
\begin{align}
 D = \psi' + \tPhi \mh\,.
\label{eq:constraintCom}
\end{align}
From the equation for $\psi'$ (\ref{eq:psi'SU_Frho}) together with \eqref{eq:tPhiCom} we obtain
\be
\psi' = - \mh \tPhi + \frac{\kappa}{6} \ma \frac{\phi'^2}{\mh} \,\tPhi
\label{eq:tPhiPsiCom}
\ee
and it immediately follows that
\begin{align}
D= \frac{\kappa}{6} \frac{\phi'^2}{\mh} \ma \tPhi\,.
\label{eq:Df}
\end{align}
We now show that $D$ satisfies
\begin{align}
\ma D' + W \ma D - \ma' D = 0
\label{eq:constraintDiff}
\end{align}
for a certain form of $W$, and with $\ma$ defined in \eqref{eq:maDef}.
Using the perturbation equation \eqref{eq:psi''SU_Frho} for $\psi''$ in the $\mf(\rho)$ case in comoving gauge (so that $\delta\phi'=0$), $D'$ reduces to
\begin{align}
D'  = \frac{N'}{N}\psi' + \mh'\,\tPhi +  \kappa \phi'^2 \left(\mfr+\frac{\rho}{2} \mfrr\right) \delta \rho = \frac{N'}{N}\psi' + \mh'\,\tPhi -  \kappa \frac{\phi'^4}{N^2} \left(\mfr+\frac{\rho}{2} \mfrr\right) \tPhi\,,
\end{align}
where we used the relation between $\delta \rho$ and $\tPhi$ given in \eqref{eq:tPhiCom}. 
It then follows that
\begin{align}
\ma D' + W \ma D - \ma' D 
= &\, \ma \tPhi\left(\mh' - \frac{N'}{N}\mh + \frac{\kappa\,\phi'^2}{6\mh }\left(\ma \frac{N'}{N}  - \ma' + W\ma \right)-\frac{\kappa \phi'^4}{ N^2}\left(\mfr + \frac{\rho}{2}\mfrr\right)\right)\\
= &\, \kappa\phi'^2\ma\, \tPhi\left(\frac{1}{6\mh }\left(\ma \frac{N'}{N}  - \ma' + W\ma -3\mh\ma\right)-\frac{\phi'^2}{ N^2}\left(\mfr + \frac{\rho}{2}\mfrr\right)\right)\\
= &\, \frac{\kappa\phi'^2\ma^2\, \tPhi}{6\mh }\left( \frac{N'}{N}  + W -3\mh\right)\,,
\end{align}
using \eqref{eq:tPhiPsiCom}, the equation for $\mh'\,$, (\ref{eq:mh'SU_Frho}), and then eliminating $\ma'$ using (\ref{eq:mfRhoSimp}).
If we now choose $W = 3 \mh - N'/ N\,$, we obtain  \eqref{eq:constraintDiff}.
Hence, as long as initial conditions are set in a regime where $D=0$ is satisfied, $\psi' + \tPhi \mh=0$ holds at all times.

We can now proceed to study the conservation laws for the gauge-invariant curvature perturbations $\zeta$ and $\mr$ defined in \eqref{gaugeinvperts} for a modified Friedmann equation of the $\mf = \mf(\rho)$ type, which is one of the main results of this paper.
As established in sec.\,\ref{sec:consZeta}, $\zeta$ is conserved whenever the adiabaticity condition  $\frac{\delta P}{P'}=\frac{\delta\rho}{\rho'}$ holds. 
If we recall that in comoving gauge,  $\delta \rho = -(\rho + P)\tPhi = \delta P\,$, 
it follows from \eqref{eq:Df} and $D=0$ that $\tPhi =0 $ and hence the adiabaticity equation is always satisfied, irrespective of the explicit form of $\pot\,$.
Therefore, in the $\mf(\rho)$ case (which, to repeat, includes both general relativity and standard LQC), $\zeta$ will always be conserved on super-horizon scales, and a single scalar matter field cannot introduce non-adiabatic perturbations. 
Furthermore, from \eqref{eq:zetaMr} it follows that $-\zeta = \mr\,$, as in general relativity.
While the intermediate steps in this argument, like the form of the constraint equation and $\delta \rho=0\,$, are gauge-dependent statements, the implications for the gauge-invariant variables $\zeta$ and $\mr$ hold in any gauge.

Another example for the $\mf(\rho)$ case can be found in the modified Friedmann equation that arises by considering Barrow entropy, as was done in \cite{barrowEntropy}. The authors give the modified Friedmann equation as $\frac{8\pi}{\kappa}\frac{ \pi^{\Delta /2} (2+ \Delta)}{(2-\Delta)} \left(\frac{\kappa}{8\pi} H^2\right)^{1-\Delta/2} = \frac{\kappa}{3} \rho$, where $N =1$,  $\Delta \in [0,1]$ is the parameter of the Barrow entropy and $\Delta \to 0$ gives the Friedmann equation of general relativity.\footnote{Here we omit the integration constant $c$, which can be identified with the cosmological constant, as it is subdominant in the early universe; and we consider the flat case $k=0$.}
 This can be rewritten as $H^2 = \frac{\kappa}{3}\rho \,\left(\frac{\kappa^2\rho}{24\pi^2}\right)^{\frac{\Delta}{2-\Delta}}\left(\frac{2-\Delta}{2+\Delta}\right)^{\frac{2}{2-\Delta}}$, i.e., $\mf = \left(\frac{\kappa^2\rho}{24\pi^2}\right)^{\frac{\Delta}{2-\Delta}}\left(\frac{2-\Delta}{2+\Delta}\right)^{\frac{2}{2-\Delta}}$.  
Since this is of the form $\mf = \mf(\rho)$, we can use the above result to conclude that also in this scenario $-\zeta = \mr\,$ as in general relativity, leading to the same conservation laws for a single scalar matter field.

\subsection{$\mf \neq \mf(\rho)$ example: GFT}
\label{sec:GFTexample}

We now turn to the more general case of $\mf \neq \mf(\rho)$, where there is no equation analogous to the diffeomorphism constraint of general relativity.
In this case, we cannot exclude non-adiabatic perturbations in general, but if we restrict ourselves to the special case of a scalar field satisfying the adiabaticity condition $\delta P = c_s^2 \delta \rho\,$, it follows that $\zeta'=0\,$, as demonstrated in sec.\,\ref{sec:consZeta}.
The quantity $\mr$ on the other hand is no longer conserved: if we insert $\delta \rho  = - (1+w)\rho\, \tPhi$ (see \eqref{eq:tPhiCom}) into the equation of motion for $\psi$ \eqref{eq:dPsi}, we find that in comoving gauge $\mr$  satisfies
\begin{align}
-\frac{\mr'}{\mh} = \tPhi + \frac{\delta \rho}{2 \rho} + \frac{\delta \mf}{2 \mf} = (1-w)\frac{\tPhi}{2} + \frac{\delta \mf}{2 \mf}\,.
\label{eq:mr'Comoving}
\end{align}
Consequently, for generalised Friedmann equations with  general $\mf$, $-\zeta = \mr$ no longer holds: while $\zeta$ remains constant on super--horizon scales, $\mr$ now has non-trivial dynamics. For a massless scalar field ($c_s^2 = w = 1$), which we consider in the following, the dynamics in $\mr$ are determined only by the expression for $\mf\,$.\\

In the remainder of this section, we investigate the dynamics of the comoving curvature perturbation $\mr$ in a GFT toy model as established in \cite{deparamcosmo}, which leads to an effective Friedmann equation specified by \eqref{eq:mfGFT}. 
The GFT framework uses a massless scalar field $\phi$ as the only matter content of the universe (or rather, it represents the dominant matter content in the bounce region one is interested in when studying quantum gravitational effects) and $\rt$ also serves as a relational matter clock.
In the special case of a massless scalar field, $\pot = 0\,$, the Klein--Gordon equation \eqref{eq:klGordon} can be solved and the expressions for the energy density and its perturbation simplify  as
\begin{align}
\phi' = \frac{\pi_\phi N}{a^3} \quad \Rightarrow \quad \rho & = \frac{\pi_\phi^2}{2 a^6}\,, \qquad
\delta \rho = 2\rho \left(\frac{\delta \pi_\phi}{\pi_\phi} + 3 \psi\right)\,, \qquad 
\rho' = -3 \mh \frac{\pi_\phi^2}{a^6} = - 6 \mh \rho\,,
\label{eq:masslesScalarField}
\end{align}
where the scalar field momentum $\pi_\phi$ is a constant of motion. 
Furthermore, for a massless scalar field, the relation between the lapse perturbation and the energy density perturbation in comoving gauge as given in \eqref{eq:tPhiCom} reduces to $\frac{\delta \rho}{\rho} = - 2 \tPhi\,$.
One also obtains 
\begin{align}
\zeta =  \frac{1}{3}\frac{\delta \pi_\phi}{\pi_\phi}\,,
\end{align}
so that the conservation of $\zeta$ follows directly from the fact that $\pi_\phi$ and its perturbation $\delta\pi_\phi$ are constants of motion. 
We first consider the evolution of $\mr \ (=\psi$ in comoving gauge) as obtained from the evolution of the GFT volume operator studied separately in each patch of the separate universe picture and then proceed to compare this to the dynamics of $\mr$ obtained by solving the generalised perturbation equation \eqref{eq:mr'Comoving}.
We limit our presentation to the main points; for details, please see app.\,\ref{app:GFT}.

\subsubsection{Evolution of $\mr$ for exact solutions in a GFT model}\label{sec:GFTquantum}

The GFT corrected Friedmann equation originates from the evolution of the expectation value of the GFT volume operator\footnote{Not to be confused with the potential of the scalar field $\pot$.} $V(\phi) := \langle \hat{V} (\phi) \rangle$ taken over a suitable class of semiclassical states with respect to the clock $\phi$.
The analytic solution for the evolution of $V(\phi)$  in a non-interacting GFT and assuming a single dominant field mode is given by \cite{deparamcosmo}
\begin{align}
V(\rt) = v_0 A e^{2 \omega \rt} + v_0 B e^{- 2 \omega \rt} - \frac{v_0}{2}\,,
\label{eq:VGFT}
\end{align}
where $A,\, B  \geq 0$ are real parameters determined by the initial conditions (and $v_0$ is a fixed constant). 
The effective Friedmann equation is then obtained from $\frac{V(\phi)'^2}{V(\phi)^2}$ (which can be related to the usual form using $V = a^3$ and rewriting $\mh$ in relational time $\phi\,$, see \eqref{eq:friedRelational}) and 
in order to obtain the correct late-time limit of this Friedmann equation the fundamental parameter $\omega$ is fixed to satisfy $\omega^2 = \frac{3}{8}\kappa\,$.

To obtain an exemplary evolution of $\mr$ directly from the solution to $V(\rt)$ as given in  \eqref{eq:VGFT}, we set up an ensemble of separate universe patches labelled by $p$, each with slightly different initial conditions $A_p, \ B_p\,$.
The bounce in each patch happens at $\phi_{p, \rm bounce} = \frac{1}{4\omega}\log \left(\frac{B_p}{A_p}\right)$, such that for generic initial conditions each patch reaches its minimum volume at a different value of $\phi\,$.
We obtain the perturbation $\psi_p$  of each patch from $V_p = (a_p)^3 = (a_{bg})^3 -3 \psi_p$ at linear order, such that 
\begin{align}
\psi_p & = \frac{1}{3}\left( 1 - \frac{V_p}{V_{bg}} \right)\,, \quad \text{where}\
V_{bg} := \frac{1}{N_{\rm patches}} \sum_p V_p\,, \label{eq:psiGFT}
\end{align}
and $N_{\rm patches}$ is the the total number of patches in the ensemble considered.
This gives an analytic expression for the perturbation $\psi_p$ (and hence $\mr_p$) of each patch.
In comoving gauge the value of $\phi$ at a given instant of relational time is (by definition) the same in each patch, and it is therefore straightforward to compare the evolution of $V_p(\phi)$ of different patches (unlike in \cite{GFTsepUniv}, where more general gauge choices were studied).

\subsubsection{Evolution of $\mr$ from separate universe perturbation equations}

We now compare the evolution of $\psi_p$ as given by \eqref{eq:VGFT} and \eqref{eq:psiGFT} to that obtained from the generalised perturbation equations \eqref{eq:dPsi} or \eqref{eq:mr'Comoving} in comoving gauge.  
We wish to establish if and for how long these generalised perturbation equations correctly capture the exact evolution of $\mr\,$.

As the matter content is given by a massless scalar field, $w=1$ in \eqref{eq:mr'Comoving} and $\mr' = -\mh \frac{\delta \mf}{2 \mf}\,$.
We fix $\mf$ as given in \eqref{eq:mfGFT}, so that
\be
\mf = 1+\frac{v_0}{a^3}+ \frac{\my}{a^6}\,, \quad \delta\mf = 3\frac{v_0}{a^3}\psi + 6 \frac{\my}{a^6}\psi + \frac{\delta \my}{a^6}\, ,
\label{eq:mfDeltaMfGFT}
\ee
and the constant of motion $\my$ is related to the coefficients in \eqref{eq:VGFT} as
\begin{align}
\my =  \frac{v_0^2}{4}-  4\,v_0^2 \, A\, B\,.
\end{align}
From the definition of $A$ and $B$ from the underlying quantum theory, it follows that $A\, B \geq \frac{1}{16}$ and hence $\my \leq 0$ (see app.\,\ref{app:GFT}).
We can consider at least two inequivalent approaches of defining the background quantity $\my = \my_{bg}$, namely $\my = \frac{v_0^2}{4} - 4\,A_{bg}\,B_{bg}$ or $\my  = \frac{1}{N_p}\sum_p \my_p\,$, where we define $A_{bg} := \frac{1}{N_{\rm patches}}\sum_p A_p$ and $B_{bg} := \frac{1}{N_{\rm patches}}\sum_p B_p\,$. 
If we average over the volumes of each patch $V_p$, which are given by  \eqref{eq:VGFT} with $A,\, B \to A_p, \, B_p\,$, we find that $V_{bg}$ is obtained by
 replacing $A,\, B$ with their background values $A,\, B \to A_{bg}, \, B_{bg}\,$
in \eqref{eq:VGFT}.
We will therefore use $\my = \frac{v_0^2}{4} - 4 A_{bg}B_{bg}$ in the following. The alternative choice would introduce nonlinear averaging effects in the evolution of $V_{bg}$ around the bounce, but these would have no impact on the qualitative statements made in the remainder of this section.
We define  $\delta \my := \my_p - \my_{bg} = -4\, v_0^2 \left( \delta A_p\, B_{bg} + \delta B_p\, A_{bg} + \delta A_p\, \delta B_p \right)\,$. 

To solve the equation of motion for $\mr$, we first 
obtain an expression for $\mh$ by solving the background Friedmann equation \eqref{eq:genFried},
which in relational time reads
\begin{align}
\mh = &   \frac{1}{3}\frac{\dd V}{\dd \rt} \frac{1}{V} \rt'
\qquad \Rightarrow \qquad \left(\frac{\dd V}{\dd \rt} \frac{1}{V}\right)^2 =  \frac{3}{2}\kappa \mf\,,
\label{eq:relFried}
\end{align}
and is solved by
\begin{align}
V(\rt) = \frac{\mc}{4}e^{ \sqrt{3 \kappa/2}\, \phi} + \left(- \my + \frac{v_0^2}{4}\right)\mc^{-1} e^{-\sqrt{3 \kappa /2}\, \phi} - \frac{v_0}{2}\,.
\label{eq:Vfried}
\end{align}

As an example, we consider again an ensemble of patches that follow \eqref{eq:VGFT} with perturbed initial conditions (different values of $A_p, \ B_p$ for each patch). These determine the value of $\my_{bg}$ and the integration constant $\mc$ is fixed by setting the initial condition from $V_{bg}$ as given by \eqref{eq:psiGFT} in the post-bounce regime.
The solution to the modified Friedmann equation  $V(\phi)$ as given  in \eqref{eq:Vfried} then agrees with the exact expression for $V_{bg}$ obtained from \eqref{eq:VGFT} and \eqref{eq:psiGFT}.\\

To now obtain the evolution of $\mr= \psi$, we solve \eqref{eq:mr'Comoving}. However, as we are concerned also with the bounce region we use the following form to avoid division by zero (since at the bounce $\mh = 0 = \mf\,$), and rewrite in relational time:
\begin{align}
2 \mh \psi' = & -\frac{\kappa}{6} \phi'^2 \delta \mf  \qquad \Rightarrow  \qquad \frac{\dd V}{\dd \rt} \frac{1}{V} \frac{\dd \psi}{\dd \rt}  =  - \frac{\kappa}{4} \delta \mf\,.
\label{eq:psi'relational}
\end{align}
Note that this  is independent of the explicit form of the lapse $N$, like the relational Friedmann equation \eqref{eq:relFried}.

\begin{figure}
     \centering
     \begin{subfigure}[t]{0.31\textwidth}
         \centering
         \includegraphics[width=\textwidth]{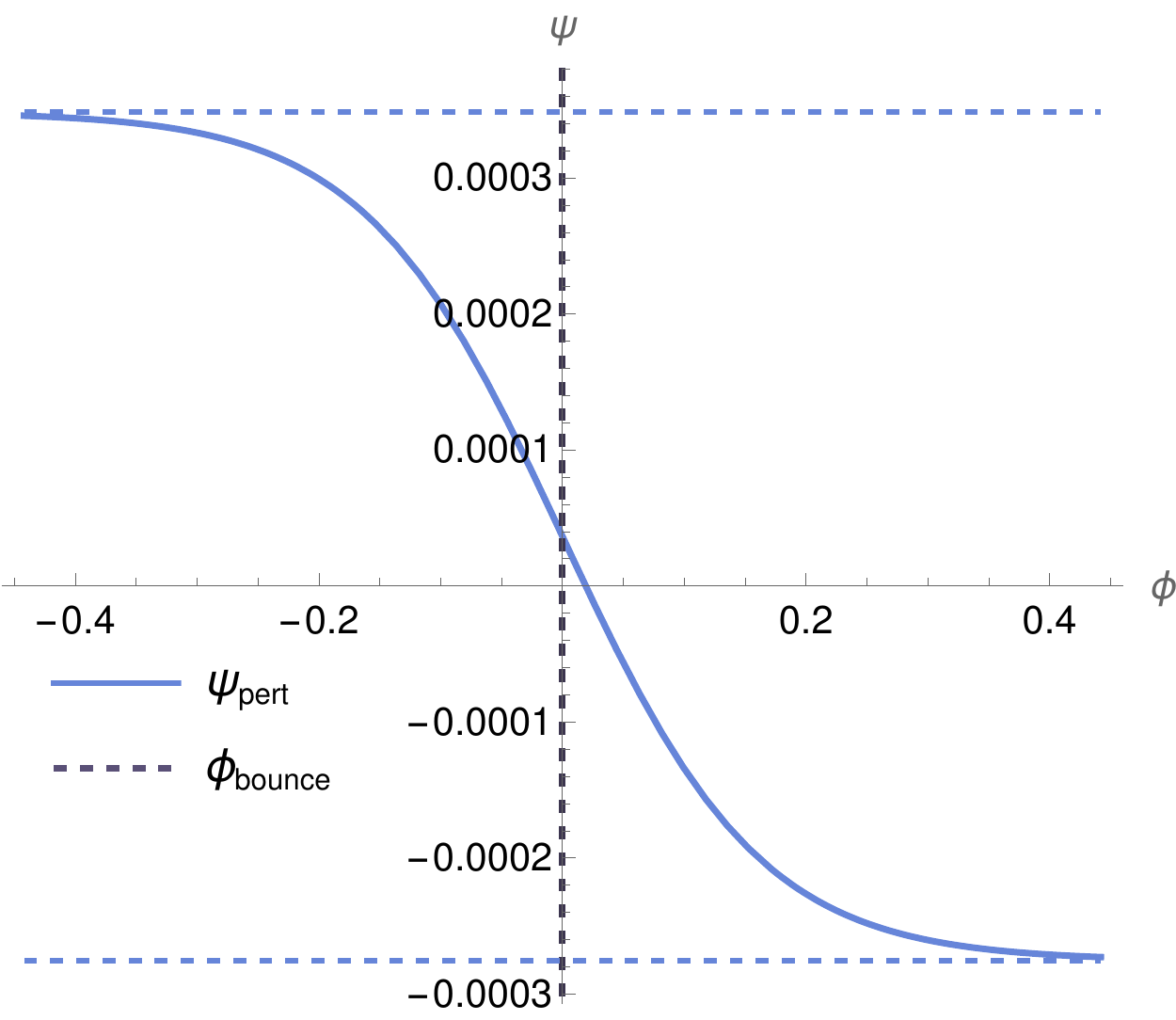}
         \caption{
Evolution of $\psi_{\rm pert}$ \eqref{eq:psiPertSol} in an exemplary patch. The horizontal dashed lines represent the asymptotic values of the solution \eqref{eq:psiAsympPertApp}.         
         }
     \end{subfigure}
     \hfill
      \begin{subfigure}[t]{0.31\textwidth}
         \centering
         \includegraphics[width=\textwidth]{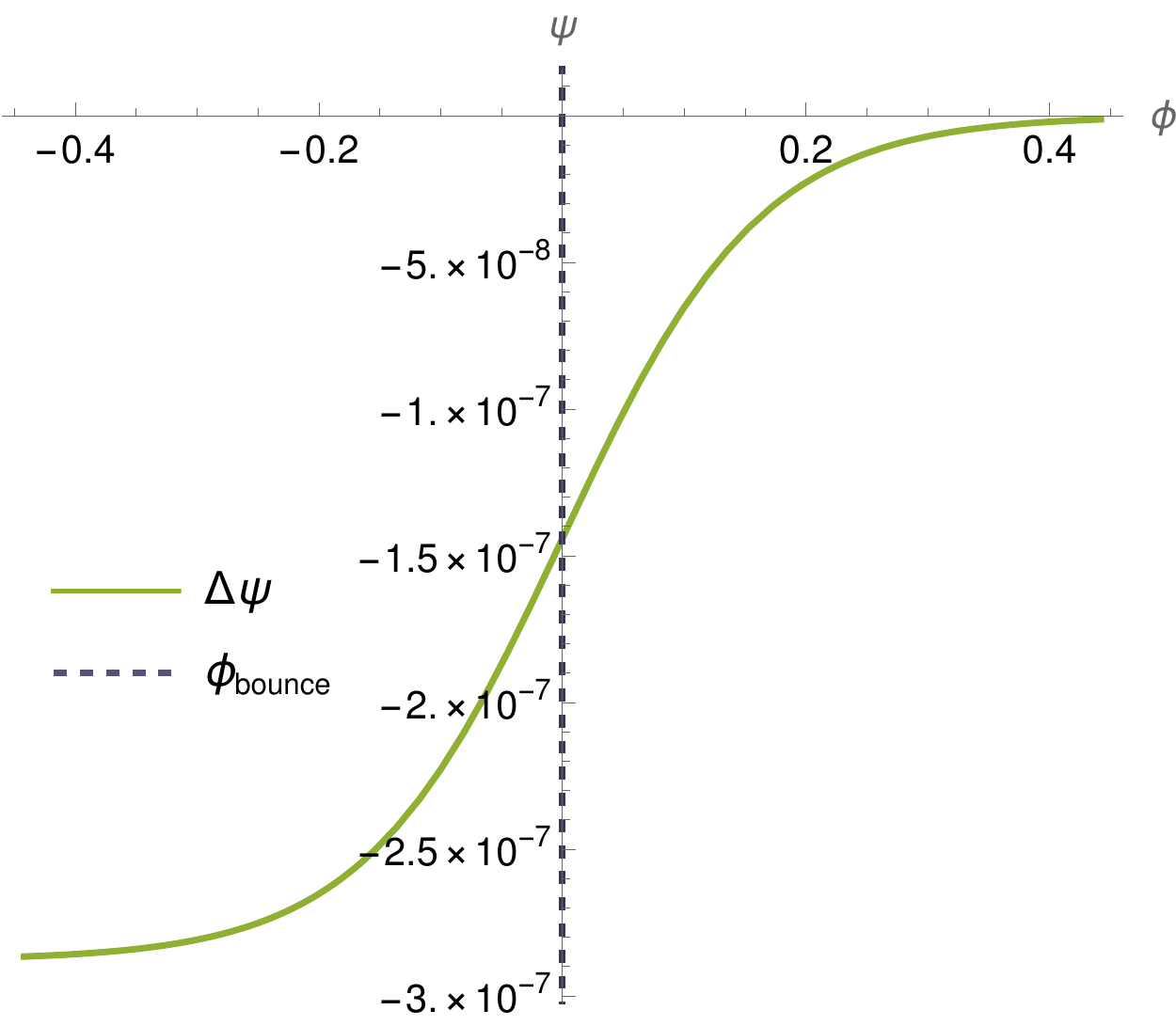}
         \caption{ The difference between the exact solution \eqref{eq:psiGFT} and the solution obtained from the perturbation equations \eqref{eq:psiPertSol}, $\Delta \psi = \psi_{\rm exact} - \psi_{\rm pert}$. 
         }
     \end{subfigure}
     \hfill
     \begin{subfigure}[t]{0.31\textwidth}
         \centering
         \includegraphics[width=\textwidth]{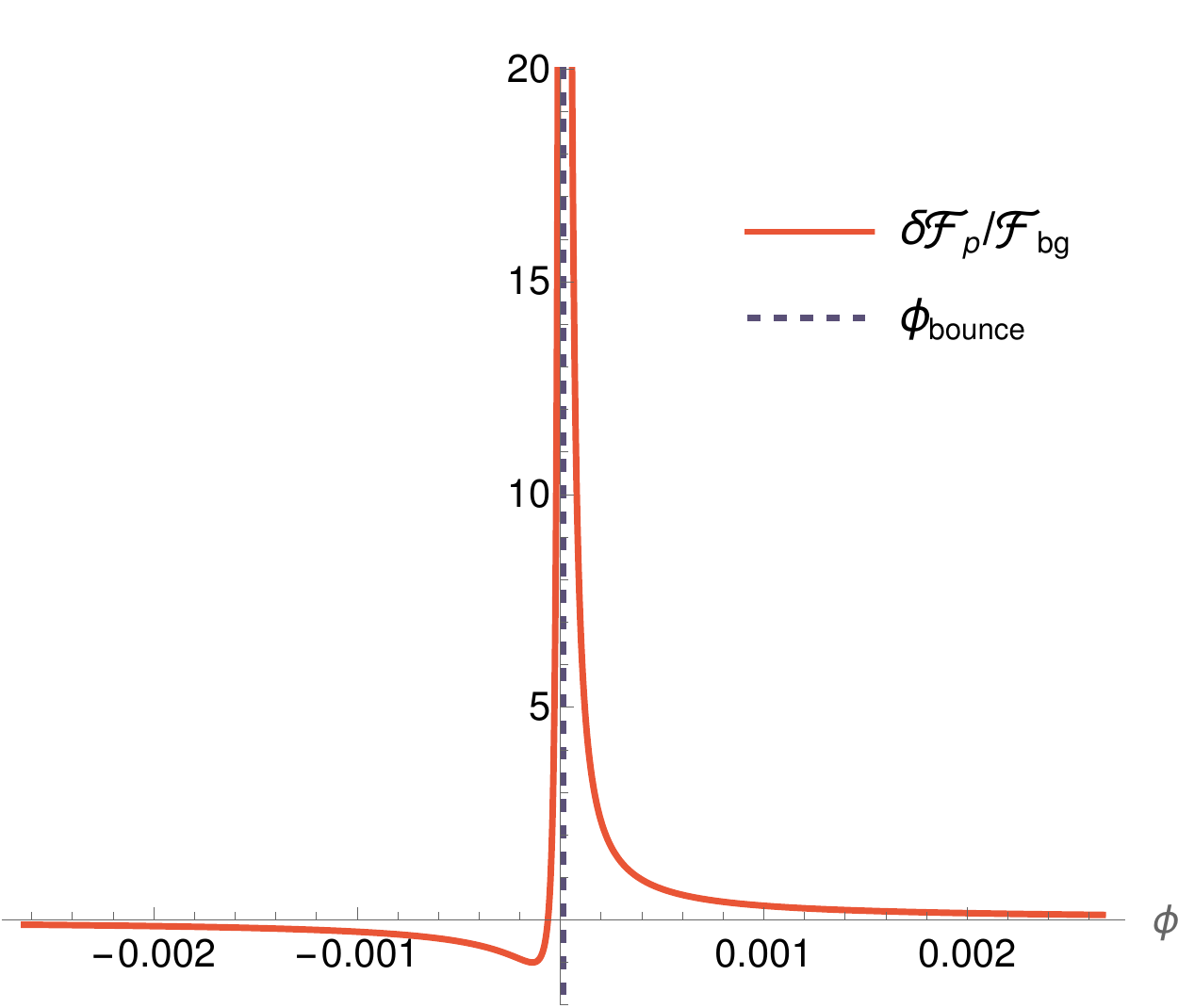}
         \caption{
         In the immediate vicinity of the bounce $\frac{\delta \mf_p}{ \mf_{bg}}$ is large, indicating a breakdown of linear perturbation theory. 
}
\label{fig:deltaMfMf}
     \end{subfigure}
        \caption{
        $\psi$ for a single patch of an ensemble with  $N_{\rm patches} = 16$ (see also fig.\,\ref{fig:exactGFT}) as given in \eqref{eq:psiPertSol} ($\psi_{\rm pert}$) compared to the exact solution \eqref{eq:psiGFT} ($\psi_{\rm exact}$). The difference between the two solutions increases in the bounce region and asymptotically approaches a constant value, but remains small throughout. Initial conditions are set in the post-bounce regime at $\phi = 4$. The asymptotic values of $\psi$ are given by \eqref{eq:psiAsympExactApp} and \eqref{eq:psiAsympPertApp}. 
While the qualitative behaviour of the plots and the conclusions we draw in the main text are independent of the specific choice of initial conditions, we quote the numerical values of parameters in the solution of $\psi_{\rm pert}$ for reference:
        $A_{bg} =200.226\,, \ B_{bg} = 200.262\,, \ \my = -160390\,, \ \mc = 800.903\,, \ \delta \my = 35.0685\,, \ \mc_\psi = 2.75576\times 10^{-4}\,, \ v_0 =1\,$. The bounce time is $\phi_{\rm bounce} = 1.47351 \times 10^{-5}$ and we set $\kappa = 8 \pi$.}
        \label{fig:q_vs_SU}
\end{figure}

A solution to \eqref{eq:psi'relational} (inserting the solution \eqref{eq:Vfried} in \eqref{eq:psi'relational}) is given by
\begin{align}
\psi =  \frac{ \mc_\psi \left(\mc^2 e^{ \sqrt{6 \kappa }\, \phi }-v_0^2 + 4 \my \right) + \frac{4}{3} \delta \my }{\left(\mc\, e^{ \sqrt{3 \kappa/2 }\, \phi }-v_0\right)^2-4 \my}\,.
\label{eq:psiPertSol}
\end{align}
The integration constant $\mc_\psi$ is fixed by setting the initial condition in the post-bounce regime from the exact solution obtained from \eqref{eq:psiGFT} for a specific patch of the ensemble.
Fig.\,\ref{fig:q_vs_SU} shows the exact evolution of $\mr$ as obtained  from \eqref{eq:psiGFT} as well as the solution to the perturbation equations \eqref{eq:psiPertSol} for a patch of an exemplary ensemble. 
If perturbations are small the exact and perturbative solution agree well. The difference in the asymptotic values in the pre-bounce regime is given by $\frac{1}{3}\frac{\delta A_p}{A_{bg}}\frac{\delta B_p}{B_{bg}}\,$.

If $\psi$ is of order $\epsilon$, the discrepancy between the asymptotic values is of order $\epsilon^2$, i.e., a quantity that is assumed negligible in linear perturbation theory. (For details, see app.\,\ref{app:asymptoticVals}.)
We would like to point out, however, that linear perturbation theory breaks down in the bounce region, since perturbations are no longer small relative to their respective background quantity: at the bounce,  $\mf =0$ but $\delta \mf\neq 0$, and hence  $\frac{\delta \mf}{\mf} \ll 1$ does not hold (fig.\,\ref{fig:deltaMfMf}). 
Note that in the special case where all patches reach their minimum volume at the same value of $\rt\,$ ($\frac{B_p}{A_p}$ is the same in all patches),  the qualitative evolution of $\psi$ differs from the example in fig.\,\ref{fig:q_vs_SU}. 
Then, even though $\psi$ is not constant around the bounce, $\mr$ takes the same value in the semiclassical regimes before and after the bounce.

We can then conclude that, despite the invalidity of linear perturbation theory in the bounce region, the generalised perturbation equations we established accurately capture   the non-trivial evolution of $\mr$ introduced by the modified Friedmann equation across the bounce if $\psi$ is sufficiently small. 
If perturbations become too large the perturbation equations reproduce the correct qualitative  behaviour of $\psi$, but lead to different values around the bounce and in the post-bounce regime.  \\

We conclude this section with two remarks: 
Even though the far pre- and post-bounce regime follow general relativistic dynamics, the relation $-\zeta = \mr$ can only hold in one of them: $\zeta$ remains conserved even in the GFT case, but $\mr$ has different asymptotic values (see fig.\,\ref{fig:q_vs_SU}). 
This can be understood by recalling that the dynamical equivalence of $-\zeta$ and $\mr$ in the $\mf(\rho)$ case follows from initially setting $D=0$ and the conservation law of $D$ (see sec.\,\ref{sec:mfRho}). If the system evolves through a period in which the conservation law for $D$ no longer holds as is the case here, this can introduce a shift between $-\zeta$ and $\mr$. Hence, in the GFT bouncing scenario where $\mr$ has non-trivial dynamics, the far pre- and post-bounce phase must be treated as independent general relativistic regimes.
Finally, the assumption that the background dynamics satisfy the same Friedmann equation as the individual locally homogeneous patches is not exact, but only holds in a perturbative regime (see app.\,\ref{app:GFT}). 
The fact that averaged quantities and their perturbations are inadequate to capture the true evolution is referred to as `the averaging problem' in standard cosmology \cite{avgProb_Zalaletdinov, avgProb_Wiltshire, avgProb_Hossenfelder}.
It can be summarised as follows: the assumption that the Universe (even at present) is homogeneous and isotropic, such that it can be described by the FLRW metric, only holds on average over large scales.
Einstein's equations are highly nonlinear and it is per se unclear whether an average of an exact solution that takes the true matter distribution of the Universe into account will match a solution obtained from perturbations around an exact FLRW universe.

\section{Relation to second order perturbation equations}\label{sec:MS}

In the previous section we studied the dynamics of the comoving curvature perturbation $\mr$ by deriving an analogue to the diffeomorphism constraint and thereby obtaining a conservation law in the $\mf(\rho)$ case, and then considering directly a solution to the first order equation in $\psi$ for a GFT model, where the matter content is given by a massless scalar field and thus takes a specific, simple form.
As established in sec.\,\ref{sec:consZeta}, $\zeta$ is conserved in the separate universe picture as long as the continuity equation holds.

Solving the first order equation \eqref{eq:dPsi} in $\psi$ directly, as we did in the GFT case, works only for specific forms of matter content, where one can eliminate all perturbation variables but one. 
In more general cases, one can combine  perturbation equations to obtain a single second order equation of motion that only refers to a single perturbation variable (and background quantities). 
This could be an equation for $\psi$ (which is equivalent to the Bardeen variable $\Psi$ in longitudinal gauge) as in \cite{0801Wands}, or the Mukhanov--Sasaki equation as in \cite{0112249}. 
Notably, these two approaches lead to different results for the evolution of $\zeta$ in the separate universe framework already in general relativity. If one obtains its evolution from a second order equation in $\psi\,$, $\zeta$ remains constant, in agreement with our considerations in sec.\,\ref{sec:consZeta}. On the other hand, if one solves the long-wavelength limit of the Mukhanov--Sasaki equation, the solution for $\zeta$ has a constant and a dynamical part, where the latter is particularly important in the contracting branch.  
We will discuss how this discrepancy could be understood as a limitation of the strict separate universe limit.

In order to relate our results to some of the literature, we summarise the  above-mentioned two second order approaches of main interest in standard cosmology and comment how they would apply to the more general types of cosmological dynamics we consider.
To simplify comparison, in this section we use conformal time ($N=a$ and we denote the Hubble parameter as $ \frac{a'}{a} = \hubb $) and longitudinal gauge ($E=B=0$ and $\psi = \tPhi\,$, where we discussed the origin of the last relation in sec.\,\ref{sec:gauge}).
In this gauge, the relevant linearised Einstein equations involving the perturbation variable $\psi$ are (see, e.g., \cite{cosmopert}) 
\begin{align}
-k^2 \psi  -3 \hubb  \left(\hubb  \psi  +\psi ' \right) & = \frac{\kappa}{2}  a ^2 \delta \rho\, , 
\label{eq:psi'GR}\\
\psi   \left(2 \hubb' +\hubb ^2\right)+3 \hubb  \psi ' +\psi '' & =\frac{\kappa}{2}  a ^2 \delta P\,. 
\label{eq:psi''GR}
\end{align}

\subsection{Second order equation for the metric perturbation $\psi$}
\label{sec:ddpsi}

One can combine the temporal \eqref{eq:psi'GR} and spatial-diagonal component \eqref{eq:psi''GR} of the Einstein equations to obtain an equation of motion for a single perturbation variable only, using that for adiabatic perturbations $\frac{\delta P}{\delta \rho} = \frac{P'}{\rho'} = c_s^2\,$. Together with the background equation $\hubb' = - \frac{1}{2}\hubb^2 (1+ 3w)\,$,
one obtains 
\begin{align}
 3 \hubb^2\left( c_s^2 - w \right)\psi +3  \hubb  (c_s^2 +1) \psi ' +\psi ''  =  -c_s^2 k^2 \psi \,  .
 \label{eq:psi''GR_k}
\end{align}
In the separate universe limit $k\rightarrow 0 \,$, one neglects the right-hand side, obtaining an equation that also follows from our earlier general separate universe equations (\ref{eq:dPsi}) and (\ref{eq:ddPsi}) in conformal time and for $\mathcal{F}=1$. In this limit, one can find the general solution $\psi(\eta) = \frac{ \hubb}{a^2} \left(\frac{3}{2} C_1 \int \dd\eta \left(a^2 \left(w+1\right)\right) + C_2\right)\,$, where $C_1\,$, $C_2$ are constants depending on $k$ in the range of wavenumbers covered by this approximation (see, e.g., \cite{Bertschinger, 0801Wands} \footnote{To see that this expression solves \eqref{eq:psi''GR_k} when $k\to0 \, $, one needs to use the background equations for $\hubb'$ and $\hubb''$ as well as the continuity equation \eqref{eq:dw}.}).

We can obtain an expression of $\zeta$ in terms of $\psi$ by replacing the energy density perturbation \eqref{eq:psi'GR} and inserting the continuity equation \eqref{continuity} in its definition \eqref{gaugeinvperts}:
\begin{align}
-\zeta=   \frac{2 k^2}{9  \hubb ^2 (w +1)} \psi +\frac{3 w  + 5}{3 (w +1)}\psi  +\frac{2 \psi ' }{3  \hubb  (w +1)}\,.
\label{eq:zetaPsi}
\end{align}
One can then derive an equation of motion for $\zeta$  which, after using \eqref{eq:psi''GR_k} and the background equations for $\hubb'$ and $w'\,$, reads 
\begin{align}
-\zeta' = \frac{2 k^2 \left( \hubb  \psi  +\psi ' \right)}{9  \hubb ^2 (w +1)}\,,
\label{eq:dZeta}
\end{align}
so that we again find $\zeta'=0$ when spatial gradients can be neglected.
In particular, one can verify explicitly that the long-wavelength solution $\psi(\eta)$ derived above results in  
\begin{align}
-\zeta(\eta) = C_1- k^2 \frac{2 C_2 + 3 C_1 \int \dd\eta\left(a ^2 (w +1)\right)}{9 a ^2 \hubb (w +1)}
\end{align}
and so again  $\zeta$ is a constant in the separate universe limit $k\rightarrow 0\,$: the second solution  $\psi\sim\frac{\hubb}{a^2}$ does not contribute.
\\

We will now investigate to what extent this approach could be illuminating for the general cosmological scenarios we consider in this paper. 
Even though it is inherent to the separate universe approach that the form of corrections  to the perturbation equations arising from spatial gradients  cannot be determined, we include a generic form of possible modifications, assuming they appear in a similar way to $k-$dependent terms in general relativity. 
Recall that in order to derive the expression for $\psi''$ in the separate universe picture \eqref{eq:ddPsi} we take the derivative of the first order equation for $\psi$ \eqref{eq:dPsi} and insert an expression for $\delta \rho'$ obtained by perturbing the continuity equation \eqref{pertcontinuity}. 
If we follow the same procedure when including inhomogeneities in $\psi'\,$, we also need to include $k-$dependent correction terms in $\delta \rho'$ to ensure consistency with general relativity: inhomogeneous changes to the dynamics of metric perturbations must lead to changes in the dynamics of the matter perturbations (whereas in the separate universe picture we assumed the matter sector remains unaltered). 
Specifically, we consider modifications to the first order equation of $\psi$ \eqref{eq:dPsi} and the perturbed continuity equation \eqref{pertcontinuity} of the form
\begin{align}
\hubb \psi' & = - \hubb^2 \left( \psi + \frac{\delta \rho}{2\rho} + \frac{\delta \mf}{2 \mf }\right) + G_k\,, \label{eq:dpsimodK}\\ 
\delta \rho' & = 3 \psi' (\rho +P) - 3 \hubb (\delta \rho + \delta P) + Z_k\, ,\label{eq:dDeltaRhoModK}
\end{align}
where in the classical limit $G_k \to -\frac{k^2}{3}\psi$ and  $Z_k \to -\frac{2 k^2}{\kappa a^2}(\psi' + \hubb \psi)\,$. 
One can then compute an expression for $\psi''$ from the derivative of \eqref{eq:dpsimodK} by inserting \eqref{eq:dDeltaRhoModK}, the continuity equation and replacing $\delta \rho$ as given by \eqref{eq:dpsimodK}, as well as making use of the background equation $\hubb' = -\frac{1}{2}\left(\hubb^2 (1+ 3w) - \hubb \frac{\mf'}{\mf}  \right)$ and the modified Friedmann equation:
\begin{align}
\begin{split}
3 \mf\hubb^2 (c_s^2-w) \psi + \left(3 \mf \hubb (c_s^2 +1) -\frac{\mf'}{2}\right)\psi' + \mf \psi'' + \left(3 \hubb^2 (c_s^2 -w) - \hubb \frac{\mf'}{\mf}\right)\frac{\delta \mf}{2} + \hubb \frac{\delta \mf'}{2} \\
 = -\frac{\kappa  a ^2 \mf^2 Z_k }{6 \hubb}+ G_k \left((3 c_s^2+1) \mf -\frac{ \mf'}{ \hubb}\right) +\frac{G_k' }{\hubb} \mf \, .
\label{eq:ddpsimodK}
\end{split}
\end{align}
In the classical limit ($\mf \to 1\,, \ \delta \mf \to 0 \,, \ \mf' \to 0,\,  \delta \mf'\to 0$ and $G_k\, , \ Z_k$ as given above)  this reduces to \eqref{eq:psi''GR_k}. 
In order to recover  \eqref{eq:dPsi}, \eqref{eq:ddPsi}  and the perturbed continuity equation \eqref{pertcontinuity} in the separate universe limit, we require $Z_k \to 0$ and $G_k \to 0$ for small $k\,$.
However, unlike for \eqref{eq:psi''GR_k}, the long-wavelength limit of \eqref{eq:ddpsimodK} cannot necessarily be solved directly to obtain a solution for $\psi\,$, as $\delta \mf$ can depend on perturbation variables other than $\psi$.

As in the general relativistic case, we can again replace $\delta \rho$ from \eqref{eq:dpsimodK} and insert the continuity and Friedmann equation in the definition of $\zeta$ \eqref{gaugeinvperts} to obtain
\begin{align}
- \zeta = -\frac{2 G_k }{3 \hubb ^2 (w +1)} +\frac{(3 w +5) \psi  }{3 (w +1)} +\frac{2 \psi ' }{3 \hubb  (w +1)}+  \frac{\delta \mf }{3 \mf  (w +1)}\,.
\end{align}
We can again compute its derivative, where we replace $\delta \rho' $ from \eqref{eq:dDeltaRhoModK}, $ \delta \rho$ from \eqref{eq:dpsimodK} and use the background equations:
\begin{align}
-\zeta' = - \frac{Z_k}{3(1+w) \rho} = - \frac{\kappa a^2 \mf Z_k}{9 \hubb^2 (1+w)} \,.
\end{align}
This reduces to \eqref{eq:dZeta} in the classical limit and vanishes when spatial gradients can be neglected.\\

In summary, following the approach used in \cite{0801Wands, Bertschinger}, we found a second order equation in $\psi$ similar to \eqref{eq:psi''GR_k}, but whether it can be written in a form that can be solved directly in the long-wavelength limit depends on the specific form of $\delta \mf\,$. 
The analysis above is consistent with the general statement from sec.\,\ref{sec:consZeta} that $\zeta'= 0$ also for a modified Friedmann equation  in the separate universe limit as long as we have adiabatic perturbations.

\subsection{Mukhanov--Sasaki equation}

In conventional cosmological perturbation  theory, one commonly works with the Mukhanov--Sasaki variable $v $ \cite{MukhanovSasaki, MukhanovSasaki_2}, which has the property of evolving like a canonically normalised scalar field in an expanding background; it appears with the standard kinetic term of a scalar field in its action, and hence can be quantised canonically as a scalar field (see, e.g., \cite{BaumannNotes}).

The dynamics of the Mukhanov--Sasaki variable are governed by the Mukhanov--Sasaki equation which (after a Fourier decomposition) reads \cite{cosmopert}
\begin{align}
v'' + c_s^2k^2v - \frac{z''}{z}v =0\,.
\label{eq:MSeq}
\end{align}
For scalar matter $v = a (\delta \phi + \frac{\phi'}{\hubb}\psi)$ and $z = a \frac{\phi'}{\hubb}$ so that $v=z\mr\,$. In general relativity and in the long-wavelength limit, one then also has $v=-z\zeta\,$.
The Mukhanov--Sasaki equation can be derived by rewriting the matter and gravity action in terms of $v$  \cite{cosmopert} or, in the separate universe approximation, through algebraic manipulation of the perturbation equations \cite{LQCsepUniv}. 
The derivation of the Mukhanov--Sasaki equation requires the constraint equation $D = \mh \tPhi +\psi' - \frac{\kappa}{2}  \phi' \delta \phi = 0\,$, which as discussed in sec.\,\ref{sec:mfRho} originates from the spatiotemporal component of the Einstein field equations and is generally not available in a separate universe framework. (But as we have seen a modified version can be derived in the $\mf(\rho)$ case.)

Another approach to obtain a solution for $\zeta$ or $\mr$ on super-horizon scales is then to solve the long-wavelength limit of \eqref{eq:MSeq}, which leads to \cite{cosmopert, 0112249}
\begin{align}
\zeta = V + S \int \frac{d \eta}{z^2}\,,
\label{eq:zetaSol}
\end{align}
where $V$ and $S$ are ($k-$dependent) constants.
Here, the dynamical part of the solution has no clear $k-$dependence that disappears in the separate universe limit. 
Imposing the long-wavelength limit of \eqref{eq:dZeta} (which is equivalent to requiring that the long-wavelength limit of the first order equation in $\psi$ is satisfied) would however recover a constant solution, by requiring $S=0\,$.
It is then unclear whether it is justified to keep the dynamical part of $\zeta$ in \eqref{eq:zetaSol}, as this amounts to neglecting $k-$dependent terms in \eqref{eq:MSeq}, but not in \eqref{eq:dZeta}. 
In some way, one then acknowledges that $\zeta' \neq 0$ for small but non-zero $k\,$, which becomes relevant in scenarios as those discussed in \cite{0112249, LQCsepUniv}, where $\zeta$ contains important information about small, but non-zero $k$ modes in the contracting phase. The authors point out that for the contracting branch in a bouncing universe,  $\zeta'$ can increase as one approaches the bounce ($-\eta \to 0$), namely in the cases where $\zeta' \sim k^2(-\eta)^{-|p|}\ (p \in \mathbb{R})$, leading to a growing mode.
To remain in the separate universe regime, one then has to assume that the wavelength of perturbations is always large enough ($k$ sufficiently small) for $\zeta'$ to remain negligible. 
However, the separate universe limit cannot be consistently applied at the bounce point, as it arises from the requirement that wavelengths are much larger than the Hubble horizon, $k\ll \hubb\,$. 
For a full treatment one would therefore need to understand the finite theory: it seems necessary to verify any statements made about the dynamics of perturbations in the separate universe limit around the bounce region against the full dynamics including gradient terms. 
In LQC there exists an effective Hamiltonian that allows to study  perturbations also outside the strict $k \to 0$ limit  and an analogue of the Mukhanov-Sasaki equation was derived  \cite{LQCanomaly}, where LQC corrections to \eqref{eq:MSeq} appear only in the $k^2$--\,term. 
This cannot be done in general for model-independent perturbation equations as we consider here. 
Furthermore, in absence of the diffeomorphism constraint it is unclear whether a Mukhanov--Sasaki like equation can be derived algebraically as was done in \cite{LQCsepUniv}\footnote{Note that in the $\mf(\rho)$ case, where $\mr' = 0$ in the separate universe framework, a second order equation $(\mr' z)' =0 $ holds independently of the choice of $z$. It is not clear how to justify a particular choice of $z$ as corresponding to the $k\rightarrow 0$ limit of an equation valid more generally.}.  
The applicability of \eqref{eq:MSeq} to a scenario with a modified Friedmann equation is therefore far from clear and an analogue needs to be established from the full dynamics for a specific model in question. 
\\

In conclusion, second order equations as are used in standard cosmology generally do not provide additional insight into the evolution of gauge-invariant perturbation variables for general theories with a modified Friedmann equation in the separate universe limit. 
They may nonetheless be useful for specific theories where additional information is available.

\section{Conclusion}\label{sec:conc}

In this article we investigated the evolution of scalar perturbations in cosmological scenarios with a modified Friedmann equation, such as those that can arise in quantum gravity. 
We focused on the gauge-invariant perturbation variables $\zeta$ and $\mr$ which are frequently studied in conventional cosmological perturbation theory, as they have a physical interpretation and are related to the power spectrum of the CMB.

Our starting point is a generic modified Friedmann equation, and the main body of our analysis is agnostic with regards to the underlying theory. 
We do however assume an unchanged continuity equation, from which it follows that $\zeta$ is conserved for long wavelengths as long as perturbations are adiabatic, independent of the gravitational dynamics.
We furthermore need to assume that the notion of gauge invariance, which ensures $\zeta$ and $\mr$ are physically meaningful variables to study, remains unchanged. 
In cases where the underlying gravitational theory admits an effective description of the modified cosmological dynamics, this could be investigated explicitly (as is the case in LQC \cite{LQCanomaly}).

We work in the separate universe framework, where the Universe is modelled as an ensemble of disconnected patches that each follow the dynamics of an FLRW universe and all spatial gradients vanish. In this framework, the perturbations are homogeneous in each patch and defined with respect to the background values of the entire ensemble.
The perturbation equations are obtained by perturbing the Friedmann equation and its derivative at linear order.
We then focus on a special case, where the modification of the Friedmann equation can be contained in a function that depends on the energy density $\rho$ only, $\mf = \mf(\rho)\,$. 
In this case one can show that a relation similar to the diffeomorphism constraint usually obtained from the spatiotemporal components of the Einstein field equations holds, which simplifies the perturbation equations. It then follows that $\mr$ is conserved for these types of models, as in general relativity. 
Similar considerations were made for  LQC in \cite{LQCsepUniv} and here we show that these results hold in general for this class of modified Friedmann equations. 
We then investigate a specific example of a Friedmann equation that does not have this property, namely the GFT Friedmann equation as established in \cite{deparamcosmo}. 
In this case, $\mr' \neq 0$ and we compare the evolution of $\mr$ across the bounce as obtained from the expectation value of the GFT volume operator to analytical solutions of the generalised perturbation equations. 
The difference between the solutions for $\psi$ obtained from the two procedures is of second order and therefore negligible for small perturbations.
Finally, we consider two common approaches in the literature that use second order equations in perturbation variables and comment how they relate to our findings. We conclude that neither of them can be used to make further general statements about the evolution of perturbations in scenarios with a general modified Friedmann equation.

In summary, we established that for a general modified Friedmann equation in the separate universe framework, the relation $\mr = -\zeta$ no longer holds, whereas it remains valid for a certain type of modification. 
As $\zeta$ remains conserved irrespective of the type of modification, the separate universe framework alone is not suitable to establish possible imprints on the CMB power spectrum from quantum gravitational effects. Inhomogeneous perturbations need to be included in an analysis to obtain alterations to the dynamics of $\zeta\,$. 
Also, this is the only way to rigorously establish how sub-horizon dynamics around the bounce influence the evolution of perturbations through the bounce.
 How and whether this can be done depends on the underlying theory that generates the modified Friedmann dynamics. 
In LQC, techniques have been established \cite{LQCanomaly}, and first investigations have also been initiated in the context of GFT \cite{GFTrelationalPert}.\\

A final comment on the definition of $\zeta$ is in order:
Here we have assumed that the definition of $\zeta$ remains unchanged also in the non-general relativistic regime. 
However, a modified Friedmann equation of the form \eqref{eq:genFried} can also be interpreted as a modification to the energy density $\rho_{\rm eff} = \rho \mf\,$, which implies a modified form of the curvature perturbation on uniform density hypersurfaces $\zeta_{\rm eff}$, as was considered in \cite{0801Wands}. The arguments for the conservation of $\zeta$ presented here would no longer apply to $\zeta_{\rm eff}\,$, since in that case $\rho_{\rm eff}' = -3\mh(\rho + P)\mf + \rho \mf'\,$. It is clear that from a Friedmann equation alone, one cannot conclude whether the modification arises in the matter sector (which then also alters the continuity equation) or is limited to gravitational dynamics. 
Such an input would originate from the theory that gives the modified Friedmann equation, and in the examples studied here (LQC and GFT) one assumes that the matter sector remains unaltered. 

\acknowledgments
The work of SG was funded
by the Royal Society through a University Research Fellowship (UF160622).

\appendix

\section{Constraint equation in Newtonian gauge}\label{app:constrNewton}

In sec.\,\ref{sec:mfRho} we showed that the diffeomorphism constraint can be retrieved in the separate universe picture in the comoving gauge. Here we carry out a similar calculation in the Newtonian gauge, which generalises what was done in \cite{LQCsepUniv} for the LQC case to any modified Friedmann equation where $\mf=\mf(\rho)\,$.
We work in conformal time $N = a$ (and denote the Hubble parameter as $\hubb$) and the longitudinal gauge $\tPhi= \psi\,$. 
Inspired by the results in \cite{LQCsepUniv}, we assume the following form of the constraint:
\begin{align}
D := \psi' + \hubb \psi -  \ma \frac{\kappa}{2} \phi' \delta \phi \,.
\end{align}
We again show that \meqref{eq:constraintDiff} holds, and hence, if initial conditions are set in a regime where $D=0$, it follows that $D'=0$ and the constraint equation holds throughout the evolution.
We use the result from sec.\,\ref{sec:mfRho} that $W = 3 \mh - N'/ N  = 2 \hubb$ in conformal time, so that we now need to show
\begin{equation}
\ma D' + 2 \hubb \ma D - \ma' D = 0\,.
\end{equation}
For this, we use the generalised equations for the $\mf(\rho)$ case as given in \eqref{eq:mh'SU_Frho}-\eqref{eq:psi''SU_Frho} and first rewrite $D'$ as
\begin{align}
D' = &  \psi'' + \hubb' \psi + \hubb \psi' -   \frac{\kappa}{2} \left(\ma'\phi' \delta \phi + \ma\phi'' \delta \phi + \ma\phi' \delta \phi'\right)  \\
= & \psi'' + \hubb'\psi -\hubb^2 \psi - \frac{\kappa \hubb}{6 \hubb} a^2 \ma\, \delta \rho -   \frac{\kappa}{2} \left(\ma'\phi' \delta \phi  -a^2 \frac{\dd\Tilde{V}}{\dd\phi} \ma \delta \phi - 2 \hubb \phi'\ma \delta \phi  + \ma\phi' \delta \phi'\right) \\
 = & \psi''  - \frac{\kappa }{6 } a^2 \ma \,\delta \rho  + \frac{\kappa}{2} \left(-\ma'\phi' \delta \phi 
 +a^2\ma\,  \delta \rho -2\phi'\ma\, \delta \phi'    + 2 \hubb \phi'\ma\, \delta \phi  \right)\,.
\end{align}
In the first step we inserted  the Klein-Gordon equation $\phi'' = -a^2\frac{\dd\Tilde{V}}{\dd\phi} - 2 \hubb \phi'$ and $\psi'$ from  \eqref{eq:psi'SU_Frho} 
and in the second made use of the background identity $\hubb' - \hubb^2 = -\frac{\kappa}{2}\phi'^2 \ma$ and used the expression for $\frac{\dd \pot}{\dd \phi} \delta \phi$  obtained from \eqref{eq:pertRhoP}. 
If we then insert $\psi''$ from \eqref{eq:psi''SU_Frho}, the expression for $D'$ can be written as
\begin{align}
D' =  \kappa \delta \rho \left( (\mf_\rho + \frac{\rho}{2} \mf_{\rho \rho}) \phi'^2   + \frac{1 }{3 } a^2 \ma \right) + \frac{\kappa}{2} \delta \phi \left(-\ma'\phi' 
   + 2 \hubb \phi'\ma  \right)\,.
\end{align}
Furthermore, making again use of  \eqref{eq:psi'SU_Frho}, we can write $D = - \frac{\kappa}{6 \hubb} a^2 \ma \delta \rho  -  \ma \frac{\kappa}{2} \phi' \delta \phi \,$, such that $2 \hubb \ma D - \ma' D $ reads 
\begin{align}
2 \hubb \ma D - \ma' D 
= & \left( - \ma^2  \frac{\kappa}{3} a^2  + \ma'\frac{\kappa}{6 \hubb} a^2 \ma  \right) \delta\rho + \left( - \hubb \ma^2 \kappa \phi' + \ma'  \ma \frac{\kappa}{2} \phi' \right) \delta \phi \,.
\end{align}
Combining the above expressions, we finally obtain
\begin{align}
\ma D' + 2 \hubb \ma D - \ma' D 
   = & \kappa \left(\mf_\rho + \frac{\rho}{2} \mf_{\rho \rho}\right) \phi'^2 \ma\,\delta \rho      +   \ma'\frac{\kappa}{6 \hubb} a^2 \ma\,   \delta\rho = 0 \,,
\end{align}
where in the last step we inserted  $\ma'$ as given in \eqref{eq:mfRhoSimp}. 

We have thus shown explicitly that a constraint equation holds also in longitudinal gauge in the case where $\mf = \mf(\rho)$. Note that in this gauge the conservation of $\mr$ is not immediately apparent from the constraint. (It can also be shown explicitly for the $\mf(\rho)$ case, but since $\mr$ is gauge-invariant, $\mr'=0$ follows from the comoving gauge.)

\section{Details of the GFT case}
\label{app:GFT}

Here we report the details of the exemplary evolution of $\psi$ studied in sec.\,\ref{sec:GFTexample}.
We first summarise briefly how the evolution of the volume of the universe $V(\phi)$ is obtained from the quantum theory and leads to an effective Friedmann equation. 
We illustrate this further by providing more specifics of the ensemble used to obtain fig.\,\ref{fig:q_vs_SU}.
We then give some more detail on how to solve the Friedmann equation to retrieve the same expression for $V(\phi)$ and include expressions for the asymptotic values of the perturbative and exact solution of $\psi\,$. 

\subsection{Exact solution}
\label{sec:exactGFTapp}

As mentioned in the main text, we use the GFT effective Friedmann equation obtained in \cite{deparamcosmo} and for the details, we refer the reader to the main paper. 
To derive the effective Friedmann equation, the authors of \cite{deparamcosmo} work in the Hamiltonian formulation of GFT \cite{GFThamiltonian}. 
We work within the free theory, i.e., neglect any interactions between GFT quanta, and furthermore restrict to a single field mode (in general the volume operator can include multiple modes).
The Hamiltonian for a single GFT mode can then be written as 
\begin{align}
\hat{H} = \frac{\hbar}{2} \omega \left((\hat{a}^\dagger)^2 + \hat{a}^2\right)\,,
\end{align}
where $\hat{a}\,, \hat{a}^\dagger$ are obtained from the GFT field operator and its momentum and are ladder operators for the Fock vacuum ($\hat{a}\ket 0 = 0$). They satisfy the bosonic commutation relations $[\hat{a},\hat{a}^\dagger]=1\,$ and dynamics are obtained by solving the Heisenberg equations of motion for $\hat{a}$ and $\hat{a}^\dagger$ ($\,\partial_\phi \hat{a} = \frac{i}{\hbar}[\hat{H}\,, \hat{a}]\,$). 
The volume operator is given by $\hat{V}(\phi) = v_0 \, \hat{a}^\dagger(\phi) \, \hat{a}(\phi)$ and its evolution with respect to the relational clock $\phi$ follows directly from the solutions to the equations of motion for $\hat{a}(\phi)$ and $\hat{a}^\dagger(\phi)\,$.

In order to infer an expression for an effective Friedmann equation, one considers the expectation value of the volume operator in a suitable semiclassical state. 
In \cite{deparamcosmo} the authors investigated different choices of coherent states and found that in the free theory, Fock coherent states can be regarded as semiclassical at late times. 
Fock coherent states $\ket \sigma = e^{\sigma \hat{a}^\dagger - \sigma^* \hat{a}}\ket 0$ are eigenstates of the annihilation operator $\hat{a}\ket\sigma = \sigma \ket \sigma$, where $\sigma \in \mathbb{C}\,$. The volume at $\phi = 0$ is then given by $\bra \sigma \hat{V}(\phi=0) \ket \sigma = |\sigma|^2$ and the evolution of the volume of the universe is \eqref{eq:VGFT}
\begin{align}
V(\rt) = \bra \sigma \hat{V}(\rt) \ket\sigma = v_0\,A e^{2 \omega \rt} + v_0\,B e^{- 2 \omega \rt} - \frac{v_0}{2}\,,
\label{eq:VGFTapp}
\end{align} 
where $A = \frac{\mRe{\sigma}^2}{2} +  \frac{\mIm{\sigma}^2}{2} +  \, \mIm{\sigma}\, \mRe{\sigma} +\frac{1}{4} $ and $B  = \frac{\mRe{\sigma}^2}{2} +  \frac{\mIm{\sigma}^2}{2} -  \, \mIm{\sigma}\, \mRe{\sigma} +\frac{1}{4}\,$.

The effective Friedmann equation for GFT is then recovered from 
\begin{align}
\left(\frac{d\bra \sigma \hat{V}(\rt) \ket\sigma}{d\phi}\right)^2 \bra\sigma \hat{V}(\rt) \ket\sigma^{-2} = 4 \omega^2 \left( 1+\frac{v_0}{\bra \sigma \hat{V}(\rt) \ket\sigma}+\frac{\my}{\bra \sigma \hat{V}(\rt) \ket\sigma^2}\right)\,,
\end{align}
where $\my =-v_0^2\left(\mRe{\sigma}^4 + \mIm{\sigma}^4 + \mRe{\sigma}^2 + \mIm{\sigma}^2 - 2 \mRe{\sigma}^2\mIm{\sigma}^2  \right)$.

\begin{figure}
     \centering
     \begin{subfigure}[t]{0.3\textwidth}
         \centering
         \includegraphics[width=\textwidth]{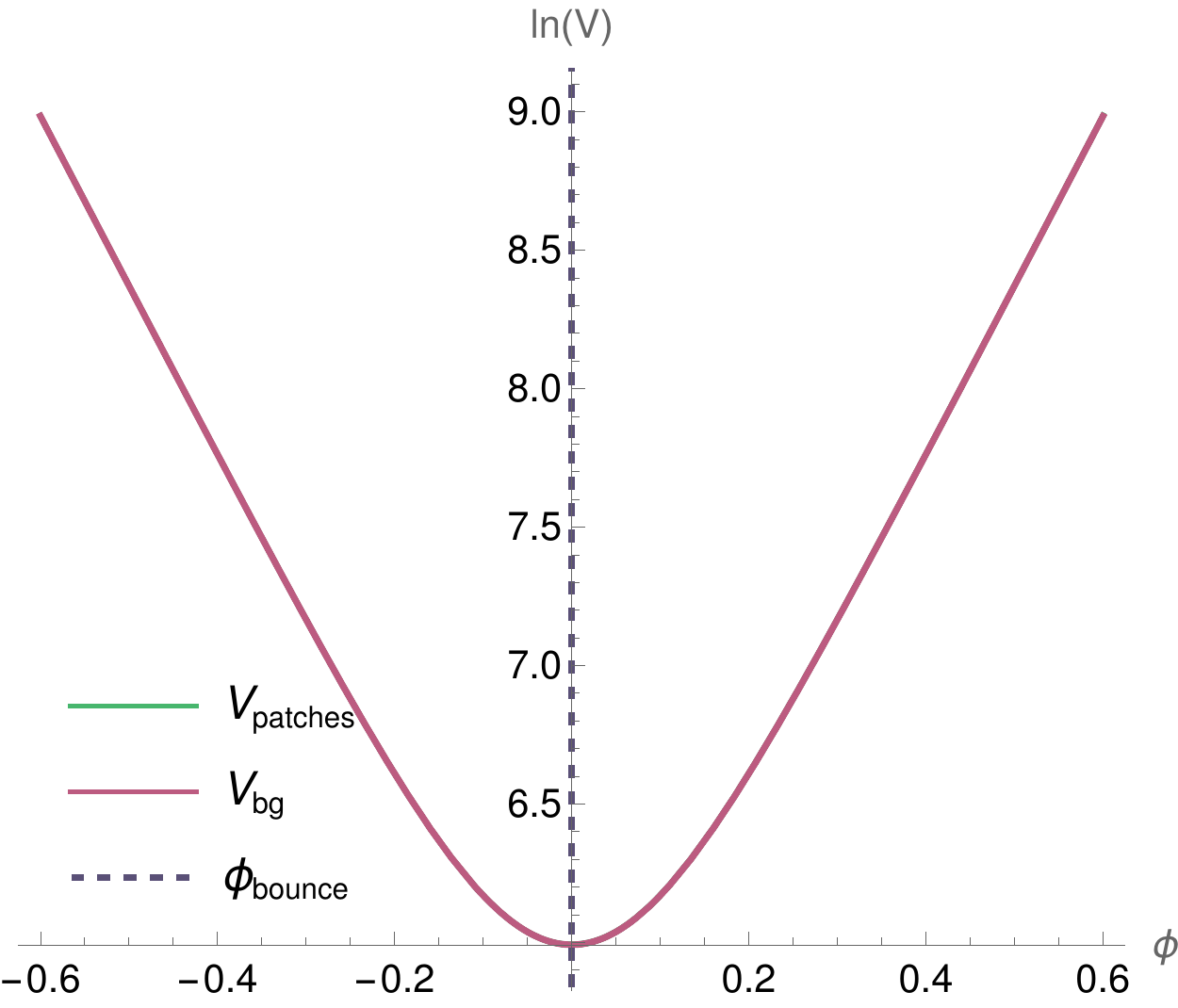}
         \caption{Evolution of the volume in each patch and the background (average) volume $V_{bg}\,$. 
         }
     \end{subfigure}
     \hfill
     \begin{subfigure}[t]{0.3\textwidth}
         \centering
         \includegraphics[width=\textwidth]{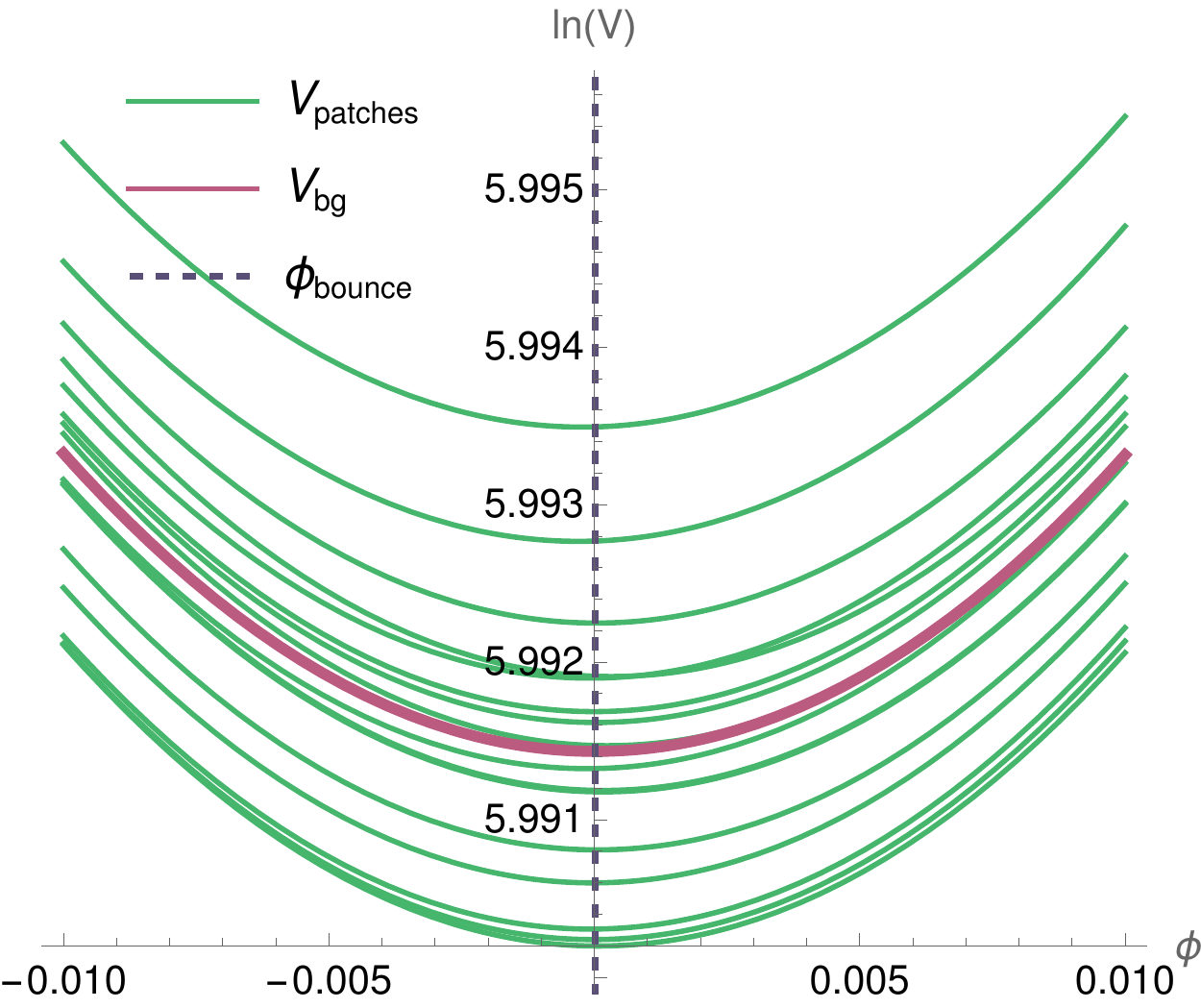}
         \caption{The evolution of the volume of each patch and $V_{bg}$ close to the bounce. The minimum volume is reached at a different value of $\phi$ in each patch.
}
     \end{subfigure}
     \hfill
     \begin{subfigure}[t]{0.3\textwidth}
         \centering
         \includegraphics[width=\textwidth]{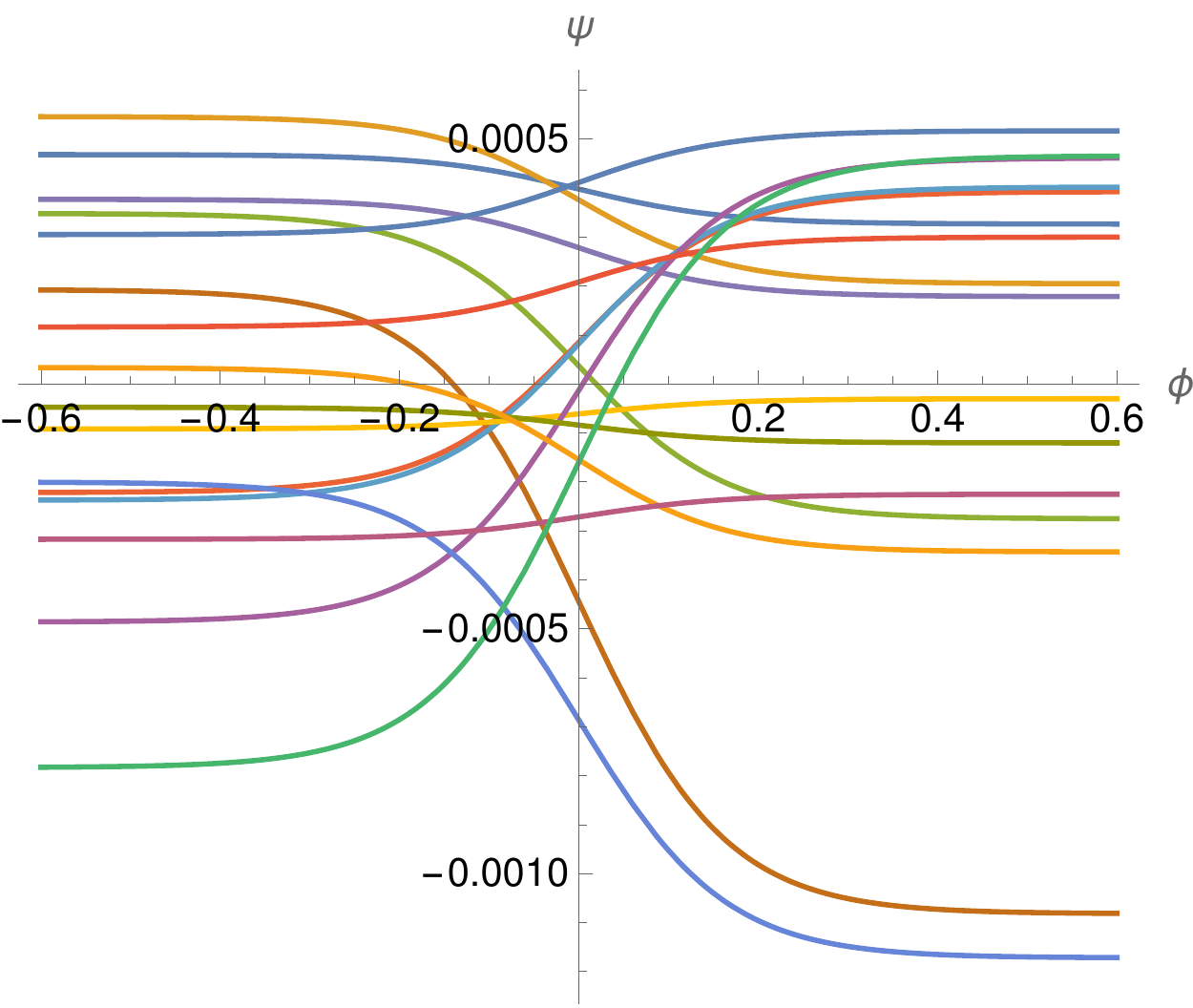}
         \caption{
Evolution of $\psi$ \eqref{eq:psiGFT} as determined by $V_p$ and $V_{bg}$ for all patches. 
}
\label{fig:psiExact}
     \end{subfigure}
        \caption{The evolution of the expectation value of the volume operator for Fock coherent states $V(\rt) = \bra{\sigma}\hat{V}(\rt)\ket{\sigma}$ and the resulting dynamics for $\psi$ in an example ensemble with  $N_{\rm patches} = 16$ and different values of $A_p\,, \ B_p$ in each patch
       \footnote{        The values in different patches (generated from a white noise process) are given in the table below. Note that the behaviour presented above is generic and the specific values are only reported for completeness. \scalebox{0.7}{
$
\begin{array}{|c|c|c|c|c|c|c|c|c|c|c|c|c|c|c|c|c|c|}
\hline
A_p & 200.03 & 200.103 & 200.391 & 199.99 & 200.119 & 200.876 & 199.984 & 200.244 & 199.948 & 200.298 & 200.046 & 200.93 & 200.432 & 200.361 & 199.946 & 199.915 \\
 B_p & 199.981 & 199.934 & 200.053 & 200.395 & 200.035 & 200.146 & 200.405 & 200.317 & 200.554 & 200.29 & 200.192 & 200.382 & 200.242 & 200.453 & 200.733 & 200.079 \\
 \hline
\end{array}
$
} }. $v_0 =1\,$.}

\label{fig:exactGFT}
\end{figure}

To obtain an exemplary evolution of $\psi\,$,  we proceed as follows:
We first choose an arbitrary number of patches for the ensemble that we wish to study. 
Then, we choose initial conditions for the parameters $A_p$ and $B_p$ in each patch, 
where random fluctuations in the initial conditions $\mRe{\sigma}$ and $\mIm{\sigma}$ are generated from a white noise process.
We then obtain $V_p(\phi)$ for each patch from the initial conditions and use this to calculate $V_{bg}$ and $\psi_p$ as defined in \eqref{eq:psiGFT}.
Fig.\,\ref{fig:exactGFT} shows the evolution of the expectation value of the volume operator for $N_{\rm patches} = 16$. The difference in evolution is strongest in the bounce region (where $\phi_{\rm bounce}$ refers to the bounce of the background, i.e., the minimum of $V_{bg}$) and each patch reaches its minimum volume at a different value of $\rt\,$. 
As a result, 
$\psi_p$ varies around the bounce, but is (approximately) constant in the far pre- and post-bounce regimes.

\subsection{Background dynamics from a modified Friedmann equation}
In order to solve the perturbation equation for $\psi$ and compare the solution to the exact evolution (fig.\,\ref{fig:psiExact}), we first need to solve the effective Friedmann equation \eqref{eq:genFried} in relational time.
To rewrite \eqref{eq:genFried}, we use $V = a^3$ to first rewrite $\mh = \frac{\dd a}{\dd \rt} \frac{1}{a} \rt' =  \frac{1}{3}\frac{\dd V}{\dd \rt} \frac{1}{V} \rt'$ and then insert the expression for $\phi' = \frac{\pi_\phi N}{V}$ and  $\rho = \frac{ \pi_\phi^2 }{2 V^2}$ \eqref{eq:masslesScalarField} into the Friedmann equation
\begin{align}
\mh^2 =  \frac{1}{9}\left(\frac{\dd V}{\dd \rt} \frac{1}{V}\right)^2 \left(\frac{\pi_\phi N}{V}\right)^2 = \frac{\kappa}{3}N^2\rho \mf 
\qquad \Rightarrow \qquad \left(\frac{\dd V}{\dd \rt} \frac{1}{V}\right)^2 =  \frac{3}{2}\kappa \mf\,.
\label{eq:friedRelational}
\end{align}
In the separate universe picture, each patch then follows $\left(\frac{\dd V_p}{\dd \rt} \frac{1}{V_p}\right)^2 =  \frac{3}{2}\kappa \mf_p$ and the perturbation of the volume in each patch $\delta V_p$ is defined by $\delta V_p = V_p - V_{bg}\,, \ V_{bg} = \frac{1}{N_{\rm patches}}\sum_p V_p\,$.
It is furthermore assumed that the average volume follows a Friedmann equation of the form $\left(\frac{d V_{bg}}{d \rt} \frac{1}{V_{bg}}\right)^2 =  \frac{3}{2}\kappa \mf_{bg}\,$.
For $\mf = \mf_{\rm GFT}$, $\mf_{bg}$ reads (we include subscripts on all quantities for clarity)
\begin{align}
 \mf_{bg} =  \frac{1}{N_p} \sum_p \mf_p = 1 + v_0 \sum_p \frac{1}{V_{bg} + \delta V_p} + \sum_p \frac{\my_{bg} + \delta \my_p}{(V_{bg} + \delta V_p)^2}  \approx  1 + \frac{v_0}{V_{bg}} + \frac{\my_{bg}}{V_{bg}^2}\, ,
\end{align}
i.e., we obtain $\mf_{bg}$ by replacing the expressions in \eqref{eq:mfGFT} with background quantities, which is an approximation that holds when perturbations are relatively small ($\frac{\delta V_p}{V_{bg}} \ll 1\,, \ \frac{\delta \my_p}{ \my_{bg}} \ll 1\,$) and linear perturbation theory is therefore applicable.
Solving the Friedmann equation  \eqref{eq:friedRelational} for $\mf = \mf_{\rm GFT}$ with $\kappa = 8 \pi G$ gives the following solution
\begin{align}
V(\phi) = \frac{\mc}{4}e^{\sqrt{3 \kappa/2 }\, \phi} + \left(- \my + \frac{v_0^2}{4}\right)\mc^{-1} e^{- \sqrt{3 \kappa/2 }\, \phi} - \frac{v_0}{2}\,.
\end{align}
The value of $\my$ is given by $\mf_{\rm GFT}$, whereas $\mc$ is an integration constant fixed by the initial condition for $V$.
If we compare to \eqref{eq:VGFT}  we find that
\begin{align}
A= \frac{\mc}{4}\,,\qquad  B = \left(- \my + \frac{v_0^2}{4}\right)\mc^{-1}\,.
\end{align}

\subsection{Asymptotic values of $\psi$ }
\label{app:asymptoticVals}

Recall that fig.\,\ref{fig:q_vs_SU} compares the exact solution obtained from \eqref{eq:psiGFT} of one patch of the ensemble (portrayed in fig.\,\ref{fig:exactGFT}) to the perturbative solution \eqref{eq:psiPertSol}.
We can 
obtain analytic expressions for the asymptotic values of both solutions.
The asymptotic values of the solution to the perturbation equations $\psi$  \eqref{eq:psiPertSol} in the far pre- and post-bounce regime, respectively, are  
\begin{align}
\psi_{\rm pre, \, pert} = \frac{4 \delta \my }{3 \left(v_0^2-4 \my \right)}-\mc_\psi\,, \qquad \psi_{\rm post, \, pert} = \mc_\psi\,.
\label{eq:psiAsympPertPreApp}
\end{align}
On the other hand, the asymptotic values of $\psi$ obtained from \eqref{eq:psiGFT} and the solution of $V$ given in \eqref{eq:VGFT} are
\begin{align}
\psi_{\rm pre,\, exact} = \frac{1}{3}\left(1- \frac{B_p}{B_{bg}}\right) = - \frac{1}{3}\frac{\delta B_p}{B_{bg}}\,, \qquad \psi_{\rm post,\, exact} = \frac{1}{3}\left(1- \frac{A_p}{A_{bg}}\right) = -\frac{1}{3}\frac{\delta A_p}{A_{bg}}\,.
\label{eq:psiAsympExactApp}
\end{align}
As we set the initial condition in the post-bounce regime, this fixes the value of $\mc_\psi = \psi_{\rm post,\, exact}\,$.
If we insert this into \eqref{eq:psiAsympPertPreApp}, we obtain
\begin{align}
\psi_{\rm pre, \, pert} = \frac{1}{3} \frac{A_p}{A_{bg}} \left( 1 - \frac{B_p}{B_{bg}} \right) = - \frac{1}{3}\frac{\delta B_p}{B_{bg}} \left( 1+ \frac{\delta A_p}{A_{bg}}\right) = \psi_{\rm pre,\, exact} + \mathcal{O}(\epsilon^2)\,.
\label{eq:psiAsympPertApp}
\end{align}
If the perturbations in $\psi$ are of order $\epsilon\,$, the difference in asymptotic values will be of order $\epsilon^2\,$, i.e., the discrepancy is a second order effect and does not affect the results of linear perturbation theory.
In order for \eqref{eq:psiAsympPertApp} to match the expression in \eqref{eq:psiAsympExactApp}, $\delta \my$ would have to be given by $\delta \my  = -4 v_0^2 (\delta A_p B_{bg} + \delta B_p A_{bg})$ instead of $\delta \my = -4 v_0^2 \left( \delta A_p B_{bg} + \delta B_p A_{bg} + \delta A_p \delta B_p \right)\,$.
The accuracy of the solution to the perturbed equations then depends on the magnitude of  $\delta A_p \,\delta B_p$ in a patch, which is determined by    the standard deviation of the white noise process we use to create perturbations in the quantum picture. 
\\

\bibliographystyle{ieeetr}

\bibliography{bib}

\end{document}